\documentclass[
  journal=pasa,
  manuscript=article-type,
  year=2022,
  volume=--,
]{cup-journal}

\pdfoutput = 1
\usepackage{amsmath}
\usepackage{microtype}
\usepackage{booktabs}
\usepackage{graphicx}	
\usepackage{amssymb}	
\usepackage{color}
\usepackage{arydshln}
\usepackage{natbib}
\usepackage{hyperref}
\hypersetup{
    colorlinks=true,
    linkcolor=blue,
    citecolor=blue,
    filecolor=blue,      
    urlcolor=blue,
    pdftitle={ABWatts VERTICO V}
    }

\usepackage{multirow}

\newcommand{\Msun}{\mathrm{M}_{\odot}}
\newcommand{\Msunpc}{\Msun\, \mathrm{pc}^{-2}}
\newcommand{\Msunyr}{\Msun\, \mathrm{yr}^{-1}}
\newcommand{\kms}{\text{km} \, \text{s}^{-1}}

\newcommand{\MHI}{M_\text{\HI}}

\newcommand{\HI}{{H}\,{\sc i}}
\newcommand{\HH}{H$_{2}$}

\newcommand{\HA}{\mathrm{H}\,{\alpha}}

\newcommand{\CO}{CO(1-0)}
\newcommand{\CCOO}{CO(2-1)}

\newcommand{\Mstar}{M_{\star}}
\newcommand{\lgMstar}{\log M_{\star}}
\newcommand{\lgMstarMsun}{\log(M_{\star}/\Msun)}

\newcommand{\lgSFRMsunyr}{\log\,\mathrm{SFR}\,[\Msunyr]}

\newcommand{\RHI}{R_\text{\HI}}

\newcommand{\SigMstar}{\Sigma_{\star}}
\newcommand{\lgSigMstar}{\log\Sigma_{\star}}
\newcommand{\lgSigMstarTR}{\langle\log\Sigma_{\star}\rangle_\text{\HI}}
\newcommand{\SigHI}{\Sigma_\text{\HI}}
\newcommand{\lgSigHI}{\log\Sigma_\text{\HI}}
\newcommand{\SigMG}{\Sigma_\text{mol}}
\newcommand{\lgSigMG}{\log\Sigma_\text{mol}}
\newcommand{\SigGas}{\Sigma_\mathrm{gas}}
\newcommand{\lgSigGas}{\log\Sigma_\mathrm{gas}}
\newcommand{\SigMGHI}{\Sigma_\text{mol} / \Sigma_\text{\HI}}
\newcommand{\lgSigMGHI}{\log(\Sigma_\text{mol} / \Sigma_\text{\HI})}

\newcommand{\lgSigMstarMsun}{\log(\Sigma_{\star}/\Msunpc)}

\newcommand{\lgSigHIMsun}{\log(\Sigma_\text{\HI}/\Msunpc)}

\newcommand{\lgSigMGMsun}{\log(\Sigma_\text{mol}/\Msunpc)}

\newcommand{\rHI}{r\HI SM}
\newcommand{\rMG}{rMGSM}

\newcommand{\cp}{\citep}
\newcommand{\ct}{\citet}
\newcommand{\fig}[1]{Fig.~\ref{fig:#1}}

\title{VERTICO V: The environmentally driven evolution of the inner cold gas discs of Virgo cluster galaxies}

\author{Adam B. Watts}
\affiliation{International Centre for Radio Astronomy Research, The University of Western Australia, Crawley, WA, Australia}
\alsoaffiliation{ARC Centre of Excellence for All-Sky Astrophysics in 3 Dimensions (ASTRO3D), Australia}
\email[Adam B. Watts]{adam.watts@uwa.edu.au}

\author{Luca Cortese}
\affiliation{International Centre for Radio Astronomy Research, The University of Western Australia, Crawley, WA, Australia}
\alsoaffiliation{ARC Centre of Excellence for All-Sky Astrophysics in 3 Dimensions (ASTRO3D), Australia}

\author{Barbara Catinella}
\affiliation{International Centre for Radio Astronomy Research, The University of Western Australia, Crawley, WA, Australia}
\alsoaffiliation{ARC Centre of Excellence for All-Sky Astrophysics in 3 Dimensions (ASTRO3D), Australia}

\author{Toby Brown}
\affiliation{Herzberg Astronomy and Astrophysics Research Centre, National Research Council of Canada, 5071 West Saanich Rd., Victoria, BC V9E 2E7, Canada}

\author{Christine D. Wilson}
\affiliation{Department of Physics \& Astronomy, McMaster University, 1280 Main Street W, Hamilton, ON, L8S 4M1, Canada}

\author{Nikki Zabel}
\affiliation{Department of Astronomy, University of Cape Town, Private Bag X3, Rondebosch 7701, South Africa}

\author{Ian D. Roberts}
\affiliation{Leiden Observatory, Leiden University, PO Box 9513, NL-2300 RA Leiden, The Netherlands}

\author{Timothy A. Davis}
\affiliation{Cardiff Hub for Astrophysics Research \&\ Technology, School of Physics \&\ Astronomy, Cardiff University, Queens Buildings, Cardiff, CF24 3AA, UK}

\author{Mallory Thorp}
\affiliation{Department of Physics \& Astronomy, University of Victoria, Finnerty Road, Victoria, BC V8P 1A1, Canada}

\author{Aeree Chung}
\affiliation{Department of Astronomy, Yonsei University, 50 Yonsei-ro, Seodaemun-gu, Seoul 03722, Republic of Korea}

\author{Adam R.H. Stevens}
\affiliation{International Centre for Radio Astronomy Research, The University of Western Australia, Crawley, WA, Australia}
\alsoaffiliation{ARC Centre of Excellence for All-Sky Astrophysics in 3 Dimensions (ASTRO3D), Australia}

\author{Sara L. Ellison}
\affiliation{Department of Physics \& Astronomy, University of Victoria, Finnerty Road, Victoria, BC V8P 1A1, Canada}

\author{Kristine Spekkens}
\affiliation{Department of Physics and Space Science, Royal Military College of Canada, PO Box 17000, Station Forces, Kingston, Ontario, Canada, K7K 7B4}

\author{Laura C. Parker}
\affiliation{Department of Physics \& Astronomy, McMaster University, 1280 Main Street W, Hamilton, ON, L8S 4M1, Canada}

\author{Yannick M. Bah{\'e}}
\affiliation{Leiden Observatory, Leiden University, PO Box 9513, NL-2300 RA Leiden, The Netherlands}

\author{Vicente Villanueva}
\affiliation{Department of Astronomy, University of Maryland, College Park, MD 20742, USA}

\author{Mar{\'i}a Jim{\'e}nez-Donaire}
\affiliation{Observatorio Astronómico Nacional (IGN), C/Alfonso XII, 3, E-28014 Madrid, Spain}
\alsoaffiliation{Centro de Desarrollos Tecnológicos, Observatorio de Yebes (IGN), 19141 Yebes, Guadalajara, Spain}

\author{Dhruv Bisaria}
\affiliation{Department of Physics, Engineering Physics, and Astronomy, Queen’s University, Kingston, ON K7L 3N6, Canada}

\author{Alessandro Boselli}
\affiliation{Aix-Marseille Université, CNRS, CNES, LAM, Marseille, France}

\author{Alberto D. Bolatto}
\affiliation{Department of Astronomy, University of Maryland, College Park, MD 20742, USA}

\author{Bumhyun Lee}
\affiliation{Korea Astronomy and Space Science Institute, 776 Daedeokdae-ro, Daejeon 34055, Republic of Korea}


\begin{document}

\begin{abstract}
The quenching of cluster satellite galaxies is inextricably linked to the {suppression} of their cold interstellar medium (ISM) by environmental mechanisms. 
While the removal of neutral atomic hydrogen (\HI) at large {radii} is well studied, how the environment impacts the remaining gas in the centres of galaxies, which are dominated by molecular gas, {is less clear}. 
Using new observations from the Virgo Environment traced in CO survey (VERTICO) and archival \HI\ data, we study the \HI\ and molecular gas within the optical discs of Virgo cluster galaxies on 1.2-kpc scales with spatially resolved scaling relations between stellar ($\SigMstar$), \HI\ ($\SigHI$), and molecular gas ($\SigMG$) {surface} densities.
Adopting \HI\ deficiency as a measure of environmental impact, we find {evidence} that, in addition to removing the \HI\ at large radii, the {cluster} {processes} also lower the average $\SigHI$ of the remaining gas {even in the central $1.2\,$kpc}. 
The impact on molecular gas is comparatively weaker than on the \HI, and we show that the lower $\SigMG$ gas is removed first. 
In the most \HI-deficient galaxies, however, we find {evidence} that  environmental {processes} {reduce} the typical $\SigMG$ of the remaining gas by nearly a factor of 3.
{We} find no evidence for environment-driven elevation of $\SigHI$ or $\SigMG$ in \HI-deficient galaxies. 
Using the ratio of $\SigMG$-to-$\SigHI$ in individual regions, we show that changes in the {ISM physical conditions, estimated using the total gas surface density and midplane hydrostatic pressure,} cannot explain the observed reduction in molecular gas content. 
Instead, we suggest that direct stripping of the molecular gas is required to explain our results. 
\end{abstract}

\section{Introduction}
Galaxies experience different evolutionary processes depending on the environment in which they live, {leading to variation in their observed properties as a function of environment}. 
In particular, galaxies in denser environments contain a systematically lower mass of cold, {gaseous} interstellar medium (ISM) than isolated systems  \cp[e.g.,][]{haynes84,giovanelli85,solanes01,fumagalli09,cortese11,catinella13,brown17,zabel19,stevens19a,stevens21}.
As this cold gas is the fuel for star formation \cp[e.g.,][]{schmidt59,kennicutt89,leroy08,bacchini19,saintonge22},  completing our understanding of the physics of galaxy evolution requires understanding how environmental mechanisms impact their cold gas content. 

Galaxy clusters are {one} of the densest environments in the Universe, where galaxies {are subject to} numerous environmental mechanisms \cp[e.g.][]{boselli06}.
Their deep gravitational potentials cause tidal stripping of satellite galaxies near the cluster center \cp{byrd90,fujita98}, and give galaxies large relative velocities \cp[$>500\,\kms$,][]{girardi93,struble99}, causing them to undergo fast gravitational interactions such as galaxy harassment \cp{moore96,mihos04}.
Additionally, clusters are permeated by a hot ($>10^7\,$K) intra-cluster medium \cp[ICM, e.g., ][]{sarazin86,mohr99,boselli06,frank13}.
Interaction with this ICM results in galaxies undergoing hydrodynamical processes such as thermal evaporation of the ISM \cp{cowie77}, turbulent stripping by viscous momentum transfer between the ICM and ISM \cp{nulsen82}, {and/}or ram-pressure stripping \cp[RPS,][]{gunn72}.
{As the removal of gas results in the future cessation of star formation, the action of these numerous environmental processes is often invoked to explain the} large fractions of quenched galaxies {hosted by clusters} \cp{dressler80,kauffmann04,peng10,woo13}. 
{While the timescale for quenching depends on the gas removal mechanism {and the circularity of the in-fall orbit \cp{font08,jung18}}, cluster satellites are consistent with their star formation being significantly reduced before or at first pericentre passage \cp[e.g., ][]{crowl08,ciesla16,boselli16a,vulcani20,werle22}, suggesting a fast gas removal process \cp[e.g.,][]{boselli06,fossati18,boselli20,upadhyay21,wright19,wright22}.}

Indeed, there is significant evidence that hydrodynamical stripping is a dominant mechanism for removing gas from cluster galaxies \cp[see the reviews by][]{cortese21,boselli22}.
{In} the canonical outside-in picture of stripping, the circumgalactic medium {is first removed} and cuts galaxies off from further accretion \cp{larson80}, before progressing to strip the ISM directly \cp[e.g.,][]{kenney04,chung07,cortese10a}. 
Further, cometary tails of stripped gas from the ISM have been observed with multi-wavelength tracers from the  X-rays to the radio continuum \cp[e.g.][]{sivanandam14,jachym14,jachym17,moretti20a,longobardi20a,roberts21,ignesti22}, and reproduced in controlled and cosmological simulations of galaxy evolution \cp[e.g.][]{tonnesen10,marasco16,yun19}.

Most of our understanding of how {environmental processes} impact the cold gas in galaxies comes from observations of the 21-cm emission line from neutral atomic hydrogen (\HI), as its large radial extent makes it a sensitive tracer of external environmental mechanisms. 
In the case of hydrodynamical stripping, galaxies exhibit clear truncation of their \HI\ distribution at a radius that depends on the strength of the {ram pressure} \cp{gunn72,cayatte90,roediger05,tonnesen09,abramson11,yoon17,reynolds21,reynolds22}.
The impact of {these environmental processes} on the molecular hydrogen (\HH) is less clear; however, truncation of the star formation in stripped galaxies \cp[e.g.,][]{koopmann04,boselli06a, cortese12b, fossati13, fossati18} indicates that the \HH\ is also affected {once stripping proceeds within the stellar disc}.
Further, the presence of star formation inside this truncation radius means that some gas can survive the stripping process. 
Whether the environment has also influenced this surviving gas, which would also affect the timescale for star formation to halt \cp[e.g.][]{boselli14,cortese21}, remains unknown.
As this surviving gas is {often} dominated by \HH\ \cp{boselli14,cortese16,loni21}, understanding the impact of {environmental processes} on the molecular ISM in galaxies is essential to completing our picture of galaxy evolution {in dense environments}.

Under most ISM conditions, however, \HH\ cannot be directly observed as it is a symmetric molecule without a permanent dipole.
Instead, most of our understanding of the molecular ISM ({from here, `molecular gas' or subscript `mol'}) comes from observing the rotational transitions of CO and its isotopologues \cp{bolatto13,saintonge17}. 
These observations have revealed that molecular gas is not affected {by environmental processes} as strongly as \HI. 
This resistance is primarily because a large fraction of molecular gas exists in giant molecular clouds (GMCs), {and it is typically more centrally concentrated where the gravitational restoring force is stronger}, making it more resistant to direct stripping than the \HI\  \cp[e.g.,][]{engargiola03,bolatto08,yamagami11,stevens21}.
Instead, {environmental processes} typically {lower} the molecular gas content in the outskirts of galaxies first  \cp{lee20,brown21,zabel22}, while significant {reduction} of the molecular gas mass of galaxies {is} only {observed} once galaxies have lost a large fraction of their total \HI\ mass \cp{fumagalli09,boselli14}. 

Although this picture is consistent with the outside-in stripping scenario, there is growing evidence that it is more complicated than just radial truncation. 
{Simulations have shown that} the multi-phase nature of the ISM allows the ICM to stream through low-density regions of the surviving gas inside the \HI\ truncation {radius}, allowing {environmental processes} to affect the cold gas across the disc \cp{tonnesen09,choi22}.
Disturbed molecular gas {morphologies} in cluster galaxies can resemble  the \HI\ \cp[e.g.,][]{lee17}, and high-resolution studies of individual objects have shown GMCs decoupled from large plumes of trailing gas \cp[e.g.,][]{cramer20}. 
Further, the mass of molecular gas  detected in the tails of stripped gas trailing behind cluster galaxies requires some level of mass-loading from the ISM \cp[e.g.,][]{sivanandam10,moretti18,lee18,jachym19}, and there is evidence for the direct stripping of cold ISM from dust observations \cp[e.g.,][]{crowl05,longobardi20a}.
Alternately, stripping of the molecular gas may not be a dominant process, and the reduction of molecular gas mass could be due to the preferential removal of \HI\ across the disc, leaving insufficient fuel to maintain efficient molecule formation \cp{fumagalli09,tonnesen09,zabel22,villanueva22}.
To further complicate the picture, there is evidence for increased molecular gas formation efficiency due to gas compression during the ICM-ISM interaction \cp[e.g.,][]{tonnesen09,moretti20,moretti20a}.
Clearly, our picture of how the environment impacts the molecular gas in galaxies is incomplete. 

In this work, we present a new analysis of the cold gas properties within the stellar components of Virgo cluster galaxies using observations from the Virgo Environment Traced in \CCOO\ \cp[VERTICO,][]{brown21} ALMA ACA\footnote{Atacama Large Millimetre/Sub-millimetre Array and Atacama Compact Array} large program.
{The} Virgo {cluster} is an excellent {target} for furthering our understanding of the environment's impact on galaxies' cold gas content.
At a distance of 16.5 Mpc \cp{mei07} it is close, allowing for detailed studies of individual galaxies, and it has been observed comprehensively across nearly the entire electromagnetic spectrum \cp{york00,chung09,davies10,urban11,boselli11,boselli18}. 
The presence of infalling sub-structures {also} means that there are populations of galaxies at various stages of infall, and thus environmental impact  \cp{gavazzi99,mei07,kim16,yoon17}. 
{Additionally,} the  VERTICO sample was selected from the VLA Imaging of Virgo galaxies in Atomic gas survey \cp[VIVA,][]{chung09} of \HI\ emission. 
Together, these data enable us to study how the {environmental processes} influence the cold gas properties of galaxies in the Virgo cluster on $\sim$1.2-kpc scales.

We present our study using the framework of spatially resolved scaling relations \cp[e.g.,][]{bigiel08, morselli20,ellison21a,ellison21,pessa22,abdurrouf22,abdurrouf22a,baker22}, in particular {those} between stellar, \HI , and molecular gas mass surface densities. 
While azimuthally averaged radial profiles are useful for increasing sensitivity to faint emission, in this work, we focus on the bright emission within the optical components of galaxies. 
Thus, comparing the physical quantities of {kpc-scale} regions within galaxies increases our statistics, and {reduces} the effects of averaging over {larger areas that include many} different physical regions. 
For analyses of how the environment impacts the radial profiles of cold gas content in VERTICO galaxies, we refer the reader to the works of \ct{brown21,zabel22}, and \ct{villanueva22}.

This paper is structured as follows. 
In \S\ref{sec:data} we describe our VERTICO sample, control sample, the homogenisation of the data, and ancillary data products. 
In \S\ref{sec:RMScomb} we investigate the resolved scaling relations with all galaxies combined and discuss some of the caveats of this approach. 
We present our main results in \S\ref{sec:RMSs}, where we study how the environment impacts the spatially resolved \HI\ and molecular gas content of galaxies across their stellar discs and within the truncation of their gas discs. 
In \S\ref{sec:MGHIr} we investigate the physical mechanisms responsible for our observations, and in \S\ref{sec:concl} we summarise our results. 

{As described in \ct{brown21},} all physical quantities used in this work were computed using a \ct{kroupa01} stellar initial mass function, a $\Lambda$CDM cosmology with $h=0.7$, $\Omega_\mathrm{M}=0.3$ and $\Omega_\Lambda=0.7$, and we adopt a constant distance for Virgo galaxies of 16.5\,Mpc \cp{mei07}.

\section{Datasets} \label{sec:data}

\begin{table*}[t]
    \centering
    \caption{Properties of the galaxies used in this work. VERTICO galaxies are contained in the top section, while the HERACLES control sample is in the bottom segment.}
    \label{tab:galprop}
    \begin{tabular}{l c c c c c c c c}
    \hline \hline
Galaxy & $D_\mathrm{L}$ [Mpc]  & PA [deg]& $i$ [deg]  & $R_{25}$ [arcmin]  & $\lgMstar$ [$\Msun$] & $\lgSFRMsunyr$ & def-\HI\ [dex] & $\lgSigMstarTR$ [$\Msunpc$] \\ 
 \hline
 &&&&&VERTICO &&&\\ \hline
NGC4254& 16.5& 243& 39 &2.68&10.52&0.70&$-0.10$&0.24 \\ 
NGC4293& 16.5& 239& 67 &2.81&10.50&$-0.27$&2.25&2.74 \\ 
NGC4298& 16.5& 132& 52 &1.62&10.11&$-0.26$&0.41&1.36 \\ 
NGC4321& 16.5& 280& 32 &3.70&10.71&0.54&0.35&1.15 \\ 
NGC4380& 16.5& 158& 61 &1.74&10.11&$-0.77$&1.13&1.69 \\ 
NGC4394$^\mathrm{a}$& 16.5& 312& 32 &1.82&10.34&$-0.79$&0.62&1.13 \\ 
NGC4450& 16.5& 170& 51 &2.62&10.70&$-0.55$&1.17&2.35 \\ 
NGC4457& 16.5& 256& 36 &1.35&10.42&$-0.49$&0.92&1.99 \\ 
NGC4501& 16.5& 320& 65 &3.46&11.00&0.43&0.58&1.64 \\ 
NGC4535& 16.5& 11& 48 &3.54&10.49&0.31&0.41&0.44 \\ 
NGC4548& 16.5& 318& 37 &2.68&10.65&$-0.28$&0.82&1.52 \\ 
NGC4567$^\mathrm{a}$& 16.5& 251& 49 &1.48&10.25&0.03&0.13&1.57 \\ 
NGC4568& 16.5& 211& 70 &2.29&10.47&0.29&0.38&1.25 \\ 
NGC4569& 16.5& 203& 69 &4.78&10.86&0.16&1.47&1.95 \\ 
NGC4579& 16.5& 273& 40 &2.94&10.92&0.08&0.95&1.88 \\ 
NGC4651& 16.5& 75& 53 &1.99&10.31&$-0.10$&$-0.30$&0.04 \\ 
NGC4654& 16.5& 300& 61 &2.45&10.26&0.31&0.12&0.79 \\ 
NGC4689& 16.5& 341& 38 &2.13&10.16&$-0.29$&0.68&1.49 \\ 
 \hline \hline
 &&&&&HERACLES &&&\\ \hline
NGC0628& 9.8& 207& 29 &4.92&10.24&0.23 & $-0.12$ &-\\
NGC3184& 13.0& 224& 13 &3.70&10.37&0.13 & $0.09$ &-\\
NGC3351& 10.0& 193& 41 &3.60&10.28&0.07 & $0.38$ &-\\
NGC3627& 10.6& 351& 59 &5.14&10.67&0.50 & $0.79$ &-\\
NGC5055& 8.9& 284& 63 &5.93&10.72&0.28 & $-0.06$ &-\\
NGC5194& 8.6& 235& 25 &3.85&10.73&0.65 & $-0.08$ &-\\
NGC5457& 7.0& 213& 31 &11.99&10.39&0.54 & $-0.13$ &-\\
NGC6946& 7.7& 250& 36 &5.70&10.50&0.79 & $-0.64$ &-\\
\hline \hline
    \end{tabular}
    \\ {\footnotesize $^\mathrm{a}$VERTICO data are 7-m array only, the rest have 7-m and Total Power array data}
    \\ {\footnotesize Columns: 1) Galaxy name; 2) Luminosity distance \cp{mei07,walter08}; 3) Position angle \cp{brown21}; 4) inclination \cp{brown21}; 5) $B$-band optical radius \cp{chung09,walter08}; 6) $\log_{10}$ integrated stellar mass \cp{leroy19}; 7) $\log_{10}$ integrated star-formation rate \cp{leroy19}; 8) \HI\ deficiency parameter \cp[][this work]{chung09}; 9) average $\lgSigMstar$ at the \HI-radius $\SigHI=1\, \Msunpc$ (this work). }
\end{table*}

\subsection{VERTICO sample} \label{subsec:sample}
{We {selected} our main galaxy sample from {the} VERTICO survey \cp{brown21}, which {obtained} \CCOO\ observations\footnote{The \CCOO\ data are publicly available at \url{https://www.canfar.net/storage/list/VERTICO}} for 51 Virgo cluster galaxies {extracted from} the VIVA survey.
We refer the reader to \ct{brown21} for a detailed description of the parent sample selection and data reduction.}
{In this work, we focus on massive ($\lgMstarMsun\geq10$) galaxies to {minimise metallicity-dependent uncertainties in tracing the molecular gas content of galaxies with \CCOO\ emission, and we enforce an optical inclination cut of $i\leq70^\circ$ to avoid large de-projection uncertainties.}
{We also removed two galaxies that met these criteria, NGC 4772 and NGC 4698, as they are passive, early-type discs that have recently accreted their \HI\ gas \cp{haynes00,cortese09,yoon17}, and thus their \HI\ and molecular gas properties do not reflect the action of the Virgo environment on their cold gas content.}
{Our final VERTICO sample consists of 18 galaxies, and the properties of the galaxies used in this work are summarised in Table \ref{tab:galprop}, along with their sources and derivations that are also described below}.}

The native {spatial} resolution {of} VERTICO data {is} $\sim8$-arcsec, and data products with 9-arcsec and 15-arcsec {resolution} were also created using the {\sc imsmooth} routine of the Common Astronomy Software Applications package \cp[CASA,][]{mcmullin07}. 
We used the {coarser} 15-arcsec data products sampled with 8-arcsec spaxels, which are closer to the spatial resolution of the \HI\ data. 
The VIVA \HI\ data used in this work were {re-imaged by author A.~Chung to circularise the synthesised beam and match the VERTICO data products as closely as possible, resulting in datacubes with 8 arcsec spaxels and synthesised beams varying from 15.2--20 arcsec, with an average of 17 arcsec.}
At the assumed 16.5 Mpc distance to Virgo, the 15/8-arcsec {beam}/spaxels of the VERTICO data correspond to a 1200/650 pc physical resolution. 

{Signal masks were created for the  \CCOO\ and \HI\ datacubes as described in \ct[][but see also \citealt{sun18} and \citealt{leroy21}]{brown21}, and moment {0} maps created for each galaxy.}
The \HI\ mass in each spaxel was calculated assuming optically thin emission \cp[e.g.,][]{meyer17},
\begin{equation}
\MHI\, [\Msun] = 2.35\times10^5 \ D_\mathrm{L}^2(1+z)^{-2}\ S_\text{\HI}
,\end{equation}
where $D_\mathrm{L}$ is the {luminosity} distance to the object {(assumed to be $16.5\,$Mpc for all VERTICO galaxies)}, {$z$ is the redshift,} and $S_\text{\HI}$ is the velocity-integrated spectral flux {density} (in Jy\,$\kms$).
Unless otherwise specified, we do not correct our \HI\ masses for the Helium abundance. 
The velocity-integrated \CCOO\ flux {density} ($S_\mathrm{CO}$) was used to calculate the total  \CCOO\ luminosity using the method of \ct{solomon05},
\begin{equation}
L_\mathrm{CO} [\mathrm{K}\, \kms\, \mathrm{pc}^{-2}] = 3.25\times10^7\, \nu^{-2}\, D_\mathrm{L}^2\, (1+z)^{-3}\, S_\mathrm{CO}
,\end{equation}
where $\nu$ is the observed frequency.
The molecular gas mass was then computed using 
\begin{equation}
M_\mathrm{mol}\, [\Msun] = \frac{\alpha_\mathrm{CO}}{R_{21}} L_\mathrm{CO}
,\end{equation}
where $R_{21}$ is the integrated flux ratio of the \CCOO\ and \CO\ lines and $\alpha_\mathrm{CO}$ {is} the molecular gas mass to \CO\ luminosity correction factor. 
We used a constant $R_{21}=0.8$, {as derived by \ct{brown21}}, and a constant $\alpha_\mathrm{CO}=4.35\,  \Msunpc$  based on  observations of the Milky Way disc by \ct{bolatto13}, which includes a 36 per cent contribution from Helium.
{The \HI\ and molecular gas mass of spaxels were converted to surface densities $\SigHI$ and $\SigMG$ in units of $\Msunpc$ using the $650$-pc physical pixel size.}

Global stellar masses were drawn from the $z=0$ Multi-wavelength Galaxy Synthesis \cp[$z$0MGS,][]{leroy19} sample, a collection of GALEX \cp[Galaxy Evolution Explorer,][]{martin05}, and WISE \cp[Wide-field Infrared Survey Explorer,][]{wright10} data for $\sim 11\,000$ galaxies with $D<50\,$Mpc and brighter than $M_\mathrm{B}=-18\,$mag.  
Stellar-mass estimates were computed from WISE band-1 observations using the WISE band-1 and -3 colour to determine the mass-to-light ratio, as described in \ct{leroy19}.
The same prescriptions were applied on a pixel-by-pixel basis to create resolved maps of stellar surface density ($\SigMstar$) at the same resolution as the \HI\ and molecular gas data. 
{Photometric} position angles {(PA)} and inclinations ($i$) for each galaxy were calculated by fitting Kron apertures to SDSS\footnote{Sloan Digital Sky Survey \cp{york00,abazajian09}} $r$-band images \cp{brown21} {and assuming an intrinsic axial ratio of $q=0.2$}, and $B$-band 25$^\mathrm{th}$ $\mathrm{mag}\, \mathrm{arcsec}^{-2}$ optical radii ($R_{25}$) were sourced from \ct{chung09}.
All surface densities used in this work were de-projected to face-on values using a factor of $\cos(i)$.

To quantify the impact of the environment on VERTICO galaxies, we {used} the type-independent \HI\ deficiency parameter (def-\HI) for each galaxy from \ct{chung09}, {which compares the observed \HI\ content to the average content in isolated galaxies with the same optical size.}
{It is defined as}
\begin{equation}
    \text{def-\HI} = \overline{\log\Sigma_\text{\HI}} - \log\Sigma_\text{\HI, obs}
\end{equation}
{where $\Sigma_\text{\HI, obs} = \text{S}_\text{\HI}/ (2\, R_{25})^2$ is the mean surface density within the optical disc.}
{Following \ct{chung09}, this definition uses $\text{S}_\text{\HI}$ in units of Jy\,$\kms$ and $R_{25}$ in arcminutes, and $\overline{\log\Sigma_\text{\HI}} = 0.37$ is the average type-independent value of  $\log(\MHI/D_{25}^2)$  from \ct[][ Table 4, 6.81]{haynes84} after a unit conversion from $\Msun\,\mathrm{kpc}^{-2}$.}
{A larger def-\HI\ value corresponds to a smaller \HI\ mass compared to isolated galaxies, and the typical scatter in the def-\HI\ distribution for  isolated galaxies is $\sim0.2-0.3$\,dex \cp[][thus automatically encapsulating the variations in the spatial distribution of \HI\ surface density observed in isolated galaxies]{haynes84,jones18a}. 
{In this work, we define `\HI-deficient' galaxies as those with def-\HI$\geq0.3 \, \mathrm{dex}$, namely, a factor of $\gtrsim$2 more \HI-poor than expected}.}


We are primarily interested in the properties of the gas within the optical disc of galaxies, so we masked each galaxy to only include pixels/spaxels within the $B$-band $R_{25}$, where the radius of each spaxel was calculated from the photometric PA and inclination. 
{While this allows us to study the impact of the environment across the discs of galaxies, we are also interested in how the environment impacts the {\emph{surviving}} cold gas disc.}
{The extent of the \HI\ gas is a good tracer of the gas truncation radius, outside of which stripping is usually assumed to be effective.}
{As our analysis uses the relationships between $\SigMstar$, $\SigHI$, and $\SigMG$, it is useful to quantify the size of the \HI\ reservoir in terms of $\SigMstar$.}

We computed radial profiles of $\lgSigHI$ and $\lgSigMstar$ using {4} arcsec annuli, {which we found gave the best consistency with the \ct{chung09} VIVA measurements.}
Defining the \HI\ truncation radius ($\RHI$) to be where the radial average $\SigHI = 1\,\Msunpc$ \cp[e.g.,][]{wang14,wang16,stevens19}, we {used} the value of  $\lgSigMstar$ at this radius to {determine} the average $\lgSigMstar$ at $\RHI$, $\lgSigMstarTR$.
{In the most \HI-rich galaxies, $\RHI$ can fall outside of the $B$-band $R_{25}$, but this does not affect any of our results as these galaxies are not the focus of our analysis.}
{Further, in all galaxies, less than 5 per cent of spaxels with $\SigMstar\geq\lgSigMstarTR$ have radius $>\RHI$, meaning trends with {higher} ({lower}) $\SigMstar$ trace trends with smaller (larger) galactocentric radius.}
{We also note here that} while the \HI\ observations cover the optical disc of each galaxy, the size of the VERTICO \CCOO\ observation footprints were based on the $\RHI$ of each target.
{Subsequently, there is some variation in the physical coverage of the molecular gas in galaxies ($0.44 - 1\times R_{25}$), and we accounted for this in our analysis of the molecular gas content.}

\begin{figure*}[t]
    \centering
    \includegraphics[width=0.49\textwidth]{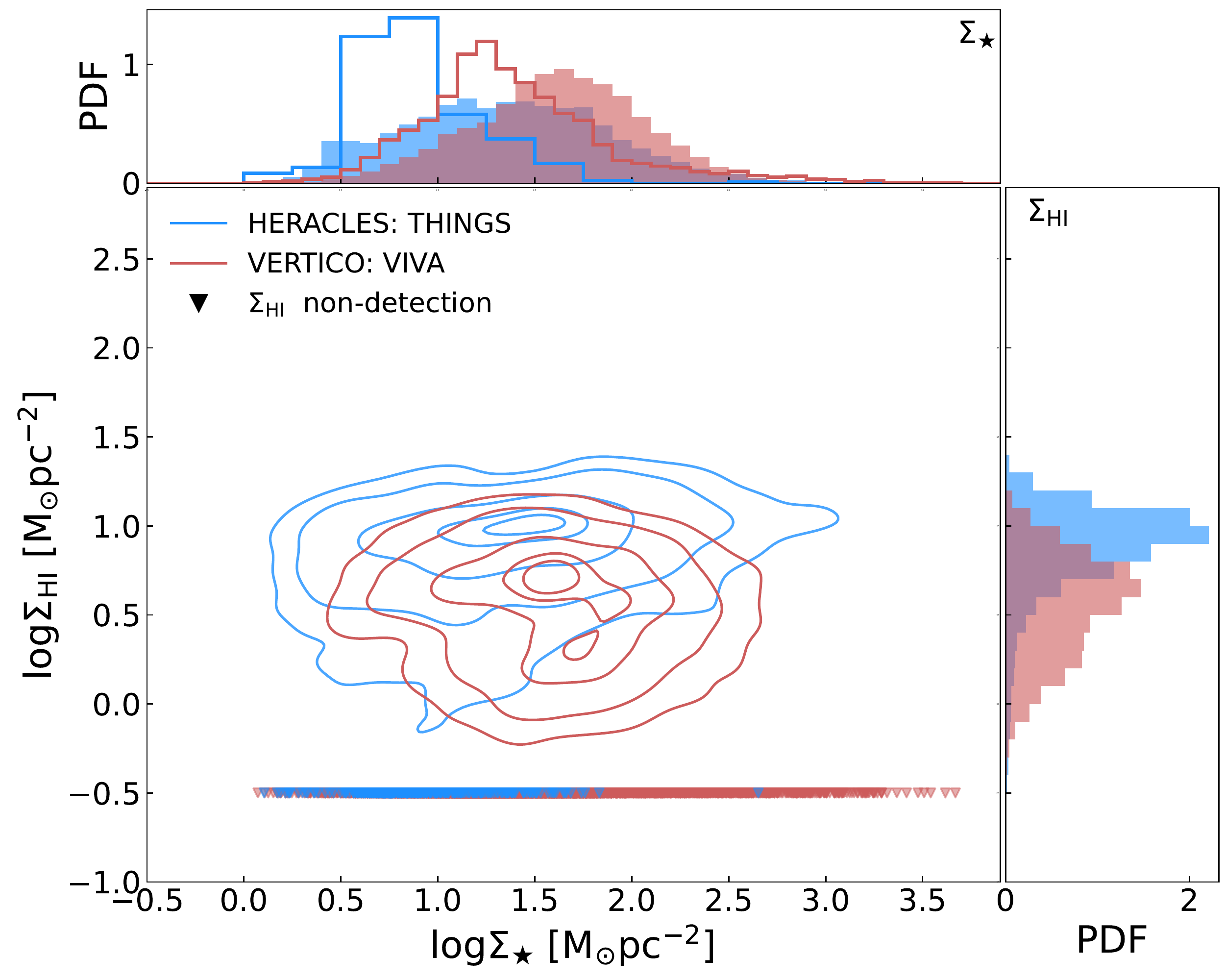}
      \includegraphics[width=0.49\textwidth]{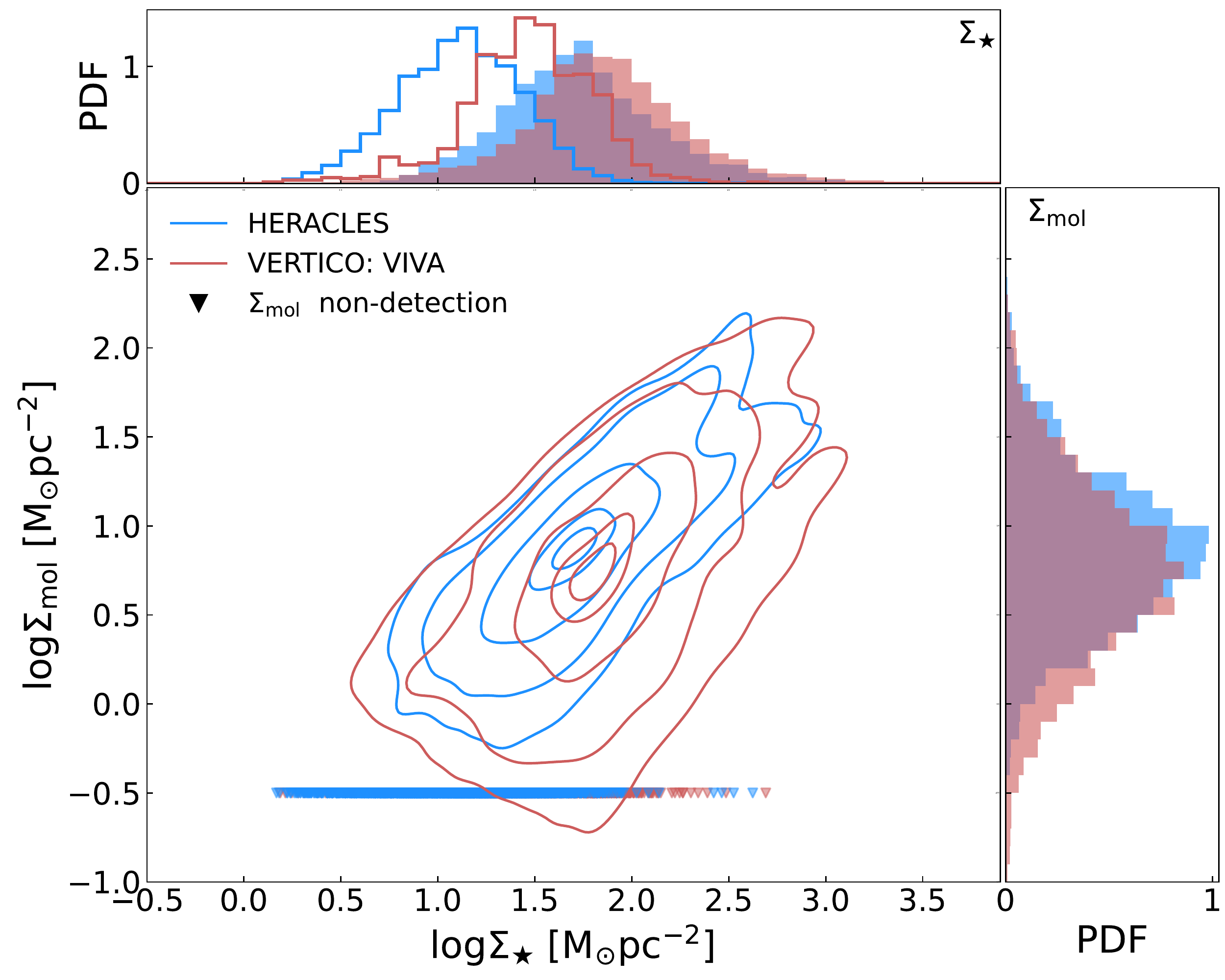}
    \caption{
    Spatially resolved $\SigHI-$ and $\SigMG-\SigMstar$ scaling relations for the combined samples. 
    The left panel shows $\lgSigHI$ vs $\lgSigMstar$, the resolved \HI--stellar mass scaling relation, and the right panel shows $\lgSigHI$ vs $\lgSigMstar$, the resolved molecular gas--stellar mass scaling relation.
    Contours for both samples enclose  5, 16, 50, 84, and 95 per cent of the \HI\ or molecular gas detected spaxels, while non-detected spaxels are shown with downward triangles at $\lgSigHIMsun=\lgSigMGMsun=-0.5$.
    Density-normalised histograms of $\lgSigMstar$, $\lgSigHI$, and $\lgSigMG$ are shown opposite their respective axes, with filled histograms representing \HI- or molecular gas-detected spaxels while open histograms represent non-detections. 
    {The HERACLES $\SigMstar$ histogram for \HI\ non-detections uses 0.25\,dex bins compared to the 0.1\,dex used for VERTICO as it has fewer points (331 compared to 5717).}
    Compared to HERACLES, the combined VERTICO sample has fewer $\SigHI$ detected spaxels, particularly at {lower} $\lgSigMstar$, and a {lower} average $\lgSigHI$. 
    Similarly, but less clearly, the combined VERTICO sample has fewer $\SigMG$ detections at {lower} $\SigMstar$, while the histogram of detected $\SigMG$ {has a larger tail toward} {lower} values. 
    }
    \label{fig:combRHIMS_RMGMS}
\end{figure*}

{While the VERTICO sample contains some \HI-normal systems, {environmental processes} can begin to impact galaxies before they become \HI-deficient.}
{Thus, we chose to draw our control sample from a different survey to ensure that we had an \HI-normal reference that was as free from environmental effects as possible}.
{We include the \HI-normal VERTICO systems in some of our figures (Figs. \ref{fig:combRHIMS_RMGMS}, \ref{fig:RHIMSpanels}, and \ref{fig:RMGMSpanels}) and discuss them briefly in \S\ref{sec:RMSs}, but the main focus of our analysis is the \HI-deficient galaxies.}

\subsection{HERACLES control sample} \label{subsec:HERACLES}
{We extracted a control sample of galaxies from the HERA\footnote{Heterodyne Receiver Array \cp{schuster04}} \CCOO\ Legacy Survey \cp[HERACLES,][]{leroy09}.}
HERACLES targeted the \CCOO\ emission line in 33 galaxies from The \HI\ Nearby Galaxy Survey \cp[THINGS,][]{walter08}, a sample of nearby ($D<15\,$Mpc), late-type galaxies spanning from dwarfs to massive spirals.
{After excluding non-detections and five galaxies that overlap with VERTICO, there are 25 HERACLES galaxies with public data \cp[for more details see][]{brown21}.}
{Physical quantities for HERACLES galaxies were  calculated using the same methods and sources as described for VERTICO,  and $B$-band $R_{25}$ were drawn from \ct{walter08}.}
{We did not compute $\lgSigMstarTR$ for HERACLES galaxies as the $\RHI$ was nearly always outside $R_{25}$.}

The HERACLES native resolution datacubes have a synthesised beam{-width} of 13 arcsec, while the THINGS data are more varied depending {on} the VLA configuration. 
Before calculating physical quantities, the HERACLES and THINGS data were smoothed using the {\sc imsmooth} routine of CASA to match the \textit{physical} {spaxel size and} resolution of the VERTICO data, and circularise the synthesised beam.
{As the distances to HERACLES sample galaxies vary, this left 11 galaxies that could be matched to a 1200-pc physical resolution in both \CCOO\ and \HI.}

To ensure that our HERACLES control sample {is} a reasonable representation of star-forming galaxies, we restricted this sample to galaxies more star-forming than  $1\sigma$ below the star-forming main-sequence using the parametrisation from the $z$0MGS sample \cp[$\sigma=0.36$, with star-formation rates (SFR) from the same catalogue,][]{leroy19}.
{After the stellar mass and inclination selection cuts, our final HERACLES control sample consists of eight galaxies, {listed in Table \ref{tab:galprop}}.}
{We compared the integrated properties of the final control sample galaxies to a subset of star-forming galaxies from the xCOLD GASS\footnote{eXtended CO Legacy Database for GASS} \cp{saintonge17} sample, selected using the same SFR threshold.}
{While the HERACLES galaxies lie on or slightly above ($<1\sigma$) the locus of the star-forming main-sequence, they span the same range in $\log\,\MHI/\Mstar$ and $\log\,M_{\mathrm{MG}}/\Mstar$ as the xCOLD GASS sample, indicating that there is no bias toward \HI- or molecule-rich systems in our control sample.}

{Last, we briefly mention the detection limits of each dataset after the homogenisation of the spatial resolutions and the masking procedures.}
{The HERACLES:THINGS data are shallower than the VERTICO:VIVA data in both \HI\ ($\SigHI\geq1-3\, \Msunpc$ within $B$-band $R_{25}$  compared to $\SigHI\geq1\, \Msunpc$) and molecular gas ($\SigMG\geq 1-3\, \Msunpc$ compared to $\SigMG\geq0.4-1\, \Msunpc$).}
{Given that we are interested in environment-driven reductions in both \HI\ and molecular gas surface densities, the fact that the VERTICO:VIVA data are deeper ensures that our findings are not due to differences in sensitivity between the two datasets.}
{The shallower nature of HERACLES means that it can have an increased fraction of non-detected spaxels near the edge of the disc, and we account for this in our analysis.}

\section{Combined $\SigHI$ --$\SigMstar$ and $\SigMG$--$\SigMstar$ relations} \label{sec:RMScomb}
\begin{figure*}[t]
    \centering
    \includegraphics[width=0.99\textwidth]{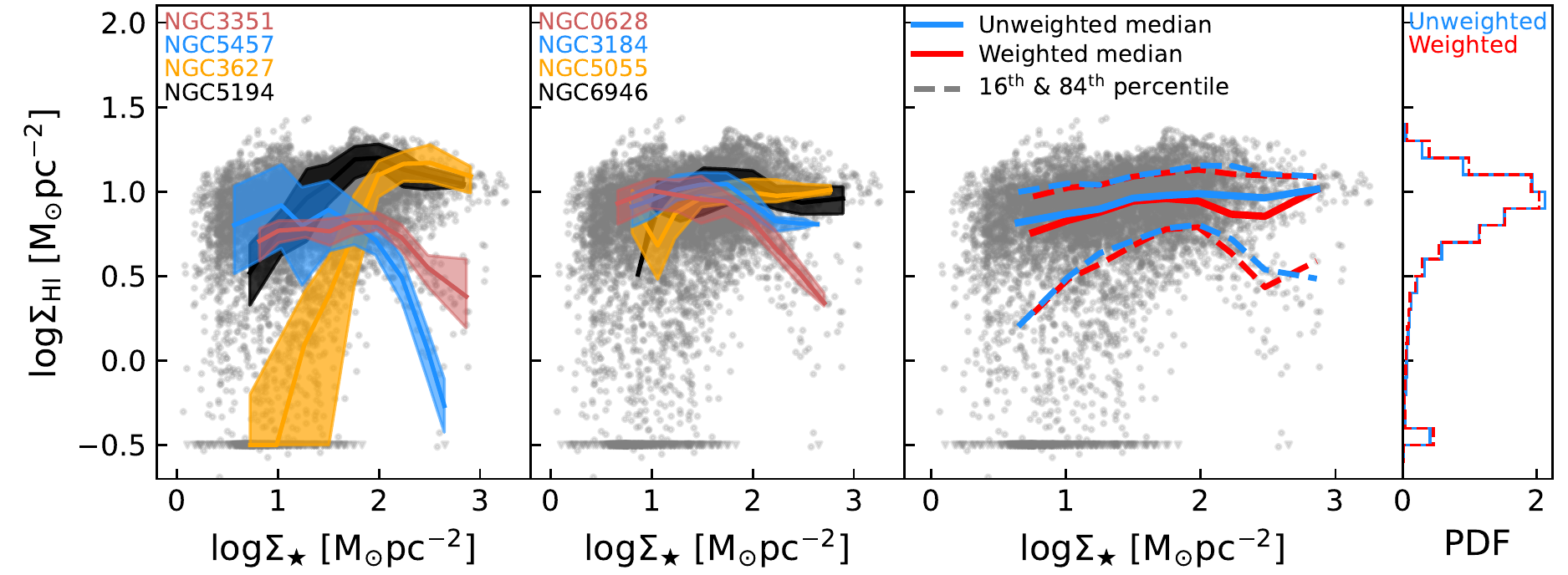}
    \includegraphics[width=0.99\textwidth]{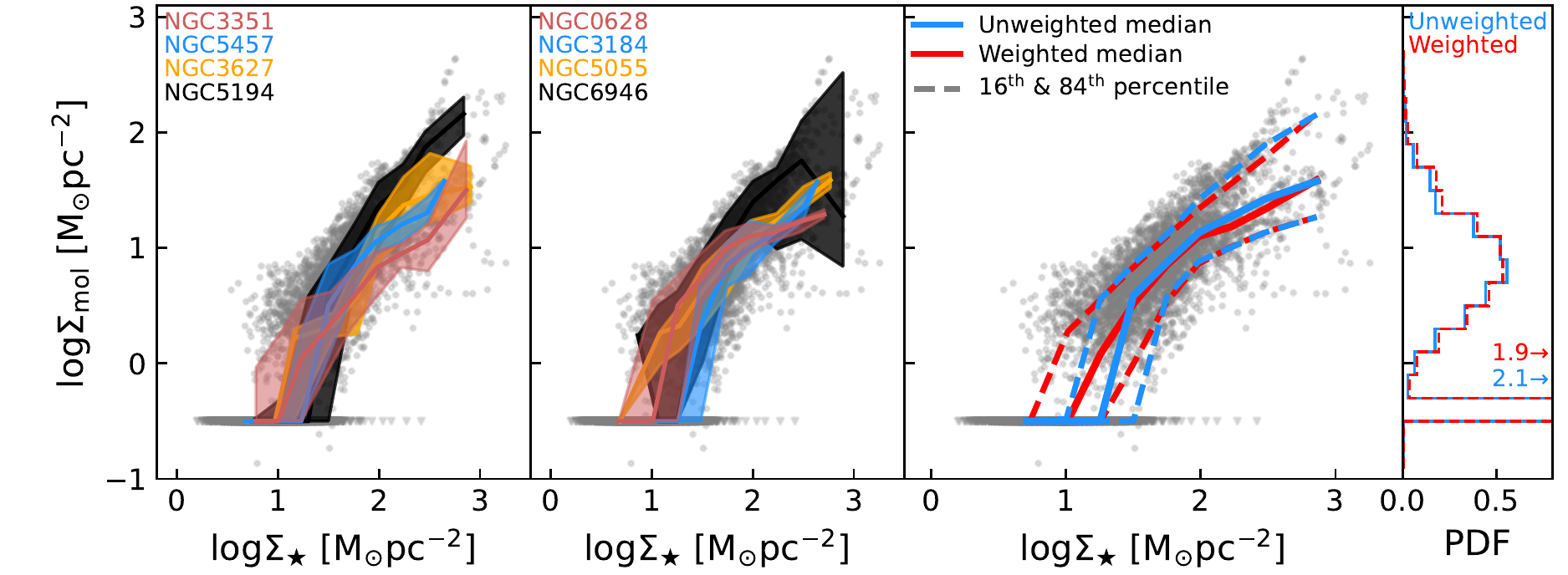}
    \caption{{{Resolved $\SigHI-$ and $\SigMG-\SigMstar$ relations for} HERACLES control galaxies. 
   Left two panels: The combined \rHI\ relation (top) and \rMG\ (bottom) are shown with grey points in the background, the median relation in bins of $\lgSigMstar$ for each galaxy is shown with thick, coloured lines, and the 16$^\mathrm{th}-84^\mathrm{th}$ percentiles are shaded in the same colour, which matches the coloured name of each galaxy in the top left corner. 
   {Galaxies are coloured from {red through to black} increasing integrated molecular gas mass and alternate between panels for better visual separation.}
   Third panel: The combined \rHI\ relation (top) and \rMG\ (bottom) are shown with grey points in the background, overlaid are the medians (solid lines) and 16$^\mathrm{th}-84^\mathrm{th}$ percentiles (dashed lines) treating all points equally (unweighted, blue) and each galaxy equally (weighted, red). 
   The rightmost panels show the density-normalised histograms for the unweighted and weighted treatments. 
   \textit{The key message of this Figure} is that galaxies lie along well-defined, individual sequences within the \rHI\ and \rMG\ relations. 
   As the location of an individual sequence is {best described} by global galaxy properties, we {choose to} weight {each galaxy equally.} 
   HERACLES galaxies do not show a large difference between the weighted and unweighted statistics, as they are restricted to the star-forming regime.}}
    \label{fig:RMSrefs}
\end{figure*}

In the left panel of \fig{combRHIMS_RMGMS} we plot $\SigHI$ against $\SigMstar$, {in other words}, the resolved \HI--stellar mass  (\rHI) relation.
{Detected spaxels for HERACLES (blue) and VERTICO (red) are shown with} contours enclosing 5, 16, 50, 84, and 95 per cent of the population, and non-detections are shown at $\lgSigHIMsun=-0.5$ with downward triangles.
Density-normalised histograms of $\lgSigMstar$ and $\lgSigHI$ are shown opposite {to} their respective axes, computed with 0.25\,dex bins. 
Filled histograms show detected spaxels, while the open histograms on the $\SigMstar$ axis show the histogram of $\lgSigMstar$ for \HI\ non-detected spaxels. 

Focusing on HERACLES galaxies, the \rHI\ relation of \HI-normal systems occupies a $\sim$0.75 dex range in $\SigHI$ across the $\sim$2.5 dex in $\SigMstar$.
\HI\ is detected across the whole $\SigMstar$ range, with non-detected spaxels having preferentially {lower} $\SigMstar$ compared to detections, also shown in the $\SigMstar$ histograms. 
{This distribution of HERACLES points} reflects the known result that \HI\ surface densities typically saturate around $\sim10\, \Msunpc$ within the stellar components of galaxies \cp{martin01,krumholz09,bigiel08,bigiel12}.
Focusing now on VERTICO, it is clear that the VERTICO combined \rHI\ relation inhabits a different region of the parameter space compared to HERACLES. 
First, there is a {decreased} number of \HI-detected spaxels across the range {of} $\SigMstar$, but preferentially at {lower} $\SigMstar$. 
This change is also visible in the $\SigMstar$ histograms where both the average VERTICO detected and non-detected spaxels are shifted to {higher} $\SigMstar$ compared to HERACLES. 
Second, the typical $\SigHI$ of VERTICO galaxies is {lower}.
{It is missing the largest $\SigHI$ values,} there is a cloud of red points at $\lgSigMstarMsun>1.5$ with $\lgSigHIMsun \sim 0-0.5$ that does not exist in HERACLES, and the $\lgSigHI$ histogram of VERTICO galaxies is broader, extending to {lower} $\SigHI$.
{{The median detected spaxel has $\lgSigHIMsun=0.58$ for VERTICO galaxies, while HERACLES has $\lgSigHIMsun=0.93$, {and a two-sample Kolmogorov-Smirnov (K-S) test rejects the null hypothesis that the two $\SigHI$ distributions are drawn from the same parent distribution with high significance ($p<10^{-4})$}.}}

Given that {the VERTICO sample spans a range of def-\HI\ values ($-0.3-2.25\, \mathrm{dex}$)} and  $\SigMstar$ increases toward the centres of galaxies, these changes are broadly consistent with galaxies undergoing ram-pressure stripping from the ICM. 
{Namely, the ISM at large radii is affected first, while lowering the average gas surface density within the central regions where the gas is more gravitationally bound requires stronger stripping that has proceeded within the stellar disc \cp{roediger07,fumagalli09,steinhauser16,yoon17}, and these two effects clearly emerge in this figure.}

How are these differences reflected in the molecular gas {surface} density of galaxies?
In the right panel of  \fig{combRHIMS_RMGMS}, we show the resolved molecular gas--stellar mass (\rMG) relation\footnote{Often called the `resolved molecular gas main sequence' in other works \cp{lin19,ellison21,pessa21}.} for the combined HERACLES and VERTICO  samples using the same colouring and non-detection labelling as the \rHI\ relation. 
Unlike $\SigHI$, $\SigMG$ shows a positive correlation with $\SigMstar$, and differences between the combined \rMG\ relation of HERACLES and VERTICO are less distinct. 
{Non-detected spaxels in both samples exist preferentially at {lower} $\SigMstar$, although the typical $\SigMstar$ of VERTICO non-detected spaxels is larger than HERACLES.}

\begin{figure*}[t!]
    \centering
    \includegraphics[width=\textwidth]{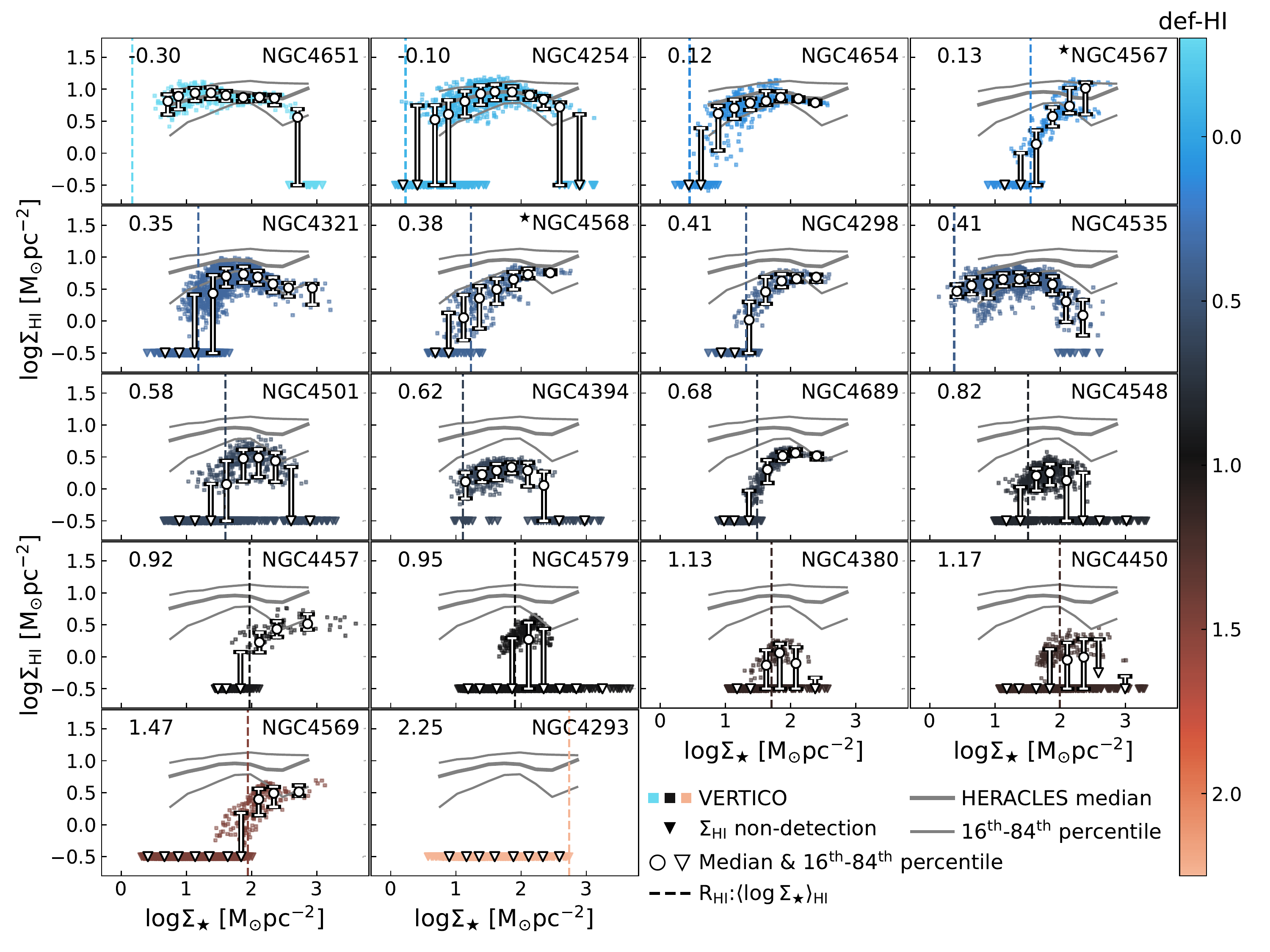}
    \caption{
    Individual \rHI\ relations for VERTICO galaxies, with the galaxy name given in the top-right {of each panel}.
    The panels are ordered in rows by def-\HI\ {(noted in the top-left corner)}, which also sets the colour of each panel, and a colour bar is displayed on the right-hand side. 
    Solid points indicate detected spaxels, while downward triangles at $\lgSigHIMsun=-0.5$ indicate non-detections. 
    The median \rHI\ for each galaxy computed in 0.25\, dex $\SigMstar$ bins is shown with large white markers, where downward triangles indicate the median is a non-detection, and {error bars show the 16$^\mathrm{th}$ and 84$^\mathrm{th}$ percentiles of the $\lgSigHI$ distribution in each bin.}
    The HERACLES control sample is shown with thick (median) and thin (16$^\mathrm{th}$ and 84$^\mathrm{th}$ percentiles), and the vertical dashed, coloured lines show $\lgSigMstarTR$, the average $\lgSigMstar$ {at the \HI\ truncation radius} ($\RHI$).
    {The two galaxies marked with stars are interacting, and their \HI\ observations are partially confused.}
    The impact of the {Virgo cluster} on the \rHI\ relation is two-fold. 
    As def-\HI\ increases there are more non-detected spaxels and $\lgSigMstarTR$ moves to larger values; and {there is an increasing reduction of the} typical $\SigHI$ within $\lgSigMstarTR$.}
    \label{fig:RHIMSpanels}
\end{figure*}

\begin{table*}[t!]
	\centering
	\caption{Weighted median $\lgSigHI$ (\fig{HI_Mstarbins}) and $\lgSigMG$ (\fig{MG_Mstarbins}) for the control sample, and the  moderate and large def-\HI\ samples, in bins of $\lgSigMstar$.
	The $\SigMstar$ range used to select spaxels are given in the first column, the second to fourth columns use all spaxels, while the remaining columns use only detected spaxels for each gas phase.  
	Uncertainties are calculated {with} {bootstrap} re-sampling {using $10^4$ iterations where entire galaxies are selected in each subsample to maintain equal weighting.}}
	\label{tab:weighted_stats}
	\begin{tabular}{lcccccc} 
		\hline \hline
		 &    \multicolumn{3}{c}{median $\lgSigHI$} &   \multicolumn{3}{c}{median detected $\lgSigHI$}  \\
		   \cmidrule(lr){2-4}\cmidrule(lr){5-7}
		 $\lgSigMstar$ bin [$\Msunpc$] & Control & def-\HI$=[0.3,0.7)$ & def-\HI$\geq0.7$ & Control & def-\HI$=[0.3,0.7)$ & def-\HI$\geq0.7$  \\
		 \hline
		 $[0,0.75)$ & 
            $0.75  \pm 0.35$  & 
            $-0.50^\mathrm{a} \pm 0.45$  & 
            $-0.50^\mathrm{a} \pm 0.00$  & 
            $0.75 \pm 0.12$  & 
            $-0.08 \pm 0.32$ &
            -\\
		 $[0.75,1.25)$ & 
            $0.83 \pm 0.09$  &
            $-0.50^\mathrm{a} \pm 0.41$ & 
            $-0.50^\mathrm{a} \pm 0.00$  & 
            $0.83 \pm 0.05$ & 
            $0.20 \pm 0.16$ & 
            $0.03 \pm 0.00$ \\
		 $[1.25,1.75)$ & 
            $0.95 \pm 0.05$  & 
            $0.33 \pm 0.15$  & 
            $-0.50^\mathrm{a} \pm 0.01$  & 
            $0.95 \pm 0.05$ & 
            $0.39 \pm 0.12$ & 
            $0.10 \pm 0.08$\\
		 $[1.75,2.25)$ & 
            $0.95 \pm 0.06$  & 
            $0.56 \pm 0.06$  & 
            $-0.14 \pm 0.26$  & 
            $0.95 \pm 0.06$ & 
            $0.57 \pm 0.06$ & 
            $0.22 \pm 0.05$ \\
		 $\geq2.25$ & 
            $0.85 \pm 0.10$  & 
            $0.49 \pm 0.15$  &
            $-0.50^\mathrm{a} \pm 0.31$  & 
            $0.85 \pm 0.10$ & 
            $0.51 \pm 0.07$ &  
            $0.32 \pm 0.06$\\  \hline \hline
		 &   \multicolumn{3}{c}{median $\lgSigMG$} &   \multicolumn{3}{c}{median detected $\lgSigMG$}  \\
		  \cmidrule(lr){2-4}\cmidrule(lr){5-7}
		 \hdashline
		 $\lgSigMstar$ bin [$\Msunpc$] & Control & def-\HI$=[0.3,0.7)$ & def-\HI$\geq0.7$ & Control & def-\HI$=[0.3,0.7)$ & def-\HI$\geq0.7$   \\ 
		 \hline
		 $[1.25,1.75)$ &
            $0.45 \pm 0.15$  &
            $0.39^\mathrm{b} \pm 0.17$  &
            $-0.50^\mathrm{a} \pm 0.26$  &
            $0.65 \pm 0.05$ &
            $0.47 \pm 0.10$ &
            $0.12 \pm 0.16$ \\
		 $[1.75,2.25)$ &
            $1.04 \pm 0.05$  &
            $0.92^\mathrm{b} \pm 0.07$  &
            $0.41 \pm 0.15$  &
            $1.04 \pm 0.06$ &
            $0.92^\mathrm{b} \pm 0.06$ &
            $0.45 \pm 0.09$ \\
		 $\geq2.25$ &
            $1.33 \pm 0.11$  &
            $1.45^\mathrm{b} \pm 0.12$  &
            $1.00 \pm 0.11$  &
            $1.34 \pm 0.11$ &
            $1.45^\mathrm{b} \pm 0.10$ &
            $1.00 \pm 0.11$ \\ 
		\hline \hline
    \end{tabular}
    \\$^\mathrm{a}$Median is a non-detection \ \ $^\mathrm{b}$Median consistent with control sample within uncertainty
\end{table*}

{The clearest difference, however, is that the VERTICO contours extend to {lower} $\SigMG$ at all $\SigMstar$, and this is true even at {higher} $\SigMstar$ where all $\SigMG$ are above the detection limit.}
{This difference is also visible in the VERTICO $\SigMG$ histogram, which extends to {lower} $\SigMG$.}
Quantitatively, the median detected VERTICO spaxel has $\lgSigMGMsun=0.73$, {lower} than the HERACLES value of $\lgSigMGMsun=0.85$, {and we can exclude that the distributions are drawn from the same parent sample ($p<10^{-4}$)}.
{Some of these differences will be due to the shallower HERACLES detection limit (causing more non-detections at {lower} $\SigMstar$) and the smaller spatial coverage of VERTICO galaxies with smaller \HI\ discs (causing an under-sampling of {lower} $\SigMstar$).}
However, they also suggest that the molecular gas is being lost preferentially at {lower} $\SigMstar$ (at least to our detection threshold), while the remaining, detectable molecular gas has lower {surface} density.
However, the magnitude of these effects appears to be much smaller, and the affected region(s) are less well defined compared to the \rHI\ relation. 

How, then, do we {extract} the comparatively weaker signal (if any) of the impact of {environmental processes} on the \rMG\ relation?
{Using a sample of galaxies from the ALMaQUEST survey \cp[ALMA-MaNGA QUEnching and STar Formation,][]{lin20} with similar $\Mstar$ and SFR ranges to our VERTICO sample, \ct{ellison21a,ellison21} showed that, to first order, the scatter within the \rMG\ relation is not determined by the properties of individual spaxels.}
Instead, galaxies form well-defined individual sequences in the parameter space, {and their} {scatter from the median relation is best described with global galaxy properties \cp[e.g., the location of an individual \rMG\ is best described by specific SFR,][]{ellison21a}}.
{We show {these individual sequences} for both the \rHI\ and \rMG\ relations of our HERACLES control sample in the left panel of \fig{RMSrefs}, {ordered from {red} to black by increasing total molecular gas mass}.}
{The distribution of HERACLES points is shown in the background, with the median relation (thick coloured lines) and $16^\mathrm{th}-84^\mathrm{th}$ percentiles (coloured, shaded regions) for each galaxy in bins of $\lgSigMstar$ overlaid.}
{Clearly, the galaxy-galaxy variation is larger than the scatter {within} individual objects, and this is particularly evident in the {higher} $\SigMstar$ regions of the \rHI\ relation, where galaxies {might} have a {central} \HI\ {{depression}}.}

{As the VERTICO sample contains both \HI-normal and \HI-deficient galaxies, combining the entire sample into one relation acts to wash out the signature of environmental effects, particularly in the molecular gas.}
{Thus, to understand how {environmental processes} impact the resolved properties of galaxies, we need to control for their global properties.}
{Further, we must also ensure that any histograms and statistics we compute are not biased toward larger galaxies and galaxies that are less affected by the environment, as these systems contain more spaxels.}
{To do this, whenever we compute a histogram such as $\lgSigHI$ or $\lgSigMG$ we weight each {spaxel} by {$1/(N_\mathrm{sp}N_\mathrm{G})$, where $N_\mathrm{sp}$ is the number of spaxels contributed by it's host galaxy and $N_\mathrm{G}$ the number of contributing galaxies.}
{Under this definition, the sum of the weights is always 1, and it ensures that each galaxy contributes to the histogram (and any subsequently computed statistics) equally.}
{All statistics presented in this paper from now on are weighted values.}

{In the middle panels of \fig{RMSrefs} we compare the weighted (red) and unweighted (blue) median (solid line), and 16$^\mathrm{th}$ and 84$^\mathrm{th}$ percentiles (dashed lines)\footnote{For $N$ data points $x_i$ with weights $w_i$, the weighted percentile $p$  is the data point $x_j$ for which the cumulative sum of the weights of the ascending-sorted $x_i$ is equal to $p$. i.e.,  $\sum_i^j w_i \leq p$ and $\sum_j^N w_i \leq (1-p)$. To compute the median, $p=0.5$.}, of the \rHI\ relation and \rMG\ relation for the HERACLES control sample.
The rightmost panels show the weighted and unweighted histograms of $\lgSigHI$ and $\lgSigMG$ across all $\SigMstar$.}
{The advantage of this weighting is clear in $\SigMG$ at $\lgSigMstarMsun\lesssim 1.5$, where the HERACLES data begin to reach their detection threshold.}
{While the unweighted statistics become non-detections, the weighted statistics better trace the relation as the typical galaxy is still detected.}
{However, to avoid biasing our statistics, we will restrict some of our quantitative $\SigMG$ comparisons to $\lgSigMstarMsun\geq1.25$, as this is the typical $\SigMstar$ where the HERACLES galaxies transition to being dominated by non-detections.}
{At these {higher} $\SigMstar$, and at all $\SigMstar$ for $\SigHI$, there {are negligible} differences between the two weighting {methods}.}
{This is expected as our control sample is defined to be  star-forming systems, {which exhibit {less} galaxy-to-galaxy variation in their spatially resolved cold gas scaling relations  \cp{ellison21a}.}}
{As we see below, this {variation is much larger} for VERTICO galaxies.}

\section{Impact of environment on the spatially resolved cold gas properties of galaxies} \label{sec:RMSs}
{Environmental effects on the \HI\ content of galaxies {are} a well-studied topic, {though many previous works have focused on the gas at large radii due to the extended nature of \HI.}
{In \S\ref{subsec:RHIMS}, we present an analysis of the \HI\ content in the central regions of galaxies using the framework of the spatially resolved scaling relations.}
{Not only does this present a new view of how {environmental processes} impact the resolved \HI\ content of galaxies, but it is also useful to}  outline the visualisations and techniques that we apply to the molecular gas in \S\ref{subsec:RMGMS}.}

\subsection{Atomic hydrogen} \label{subsec:RHIMS}

In \fig{RHIMSpanels}, we present the \rHI\ relations of VERTICO galaxies in individual panels. 
\HI-detected spaxels are shown with coloured points and non-detections with downward triangles. 
The median $\lgSigHI$ in $0.25$\, dex bins of $\SigMstar$ (which are combined with adjacent bins if they contain less than 10 points) is shown with white, bordered symbols, where a  circle denotes a median detection and a downward triangle denotes that the median is a non-detection. 
The median (thick, grey line) and 16$^\mathrm{th}$ and 84$^\mathrm{th}$ percentiles (thin, grey lines) of the HERACLES control sample are shown in the background, and the average $\lgSigMstar$ measured at  $\RHI$ ($\lgSigMstarTR$, as calculated in \S\ref{subsec:sample}) is shown with a vertical dashed line. 
Panels are ordered in rows from left to right {by} increasing def-\HI\ {(the numerical value is displayed in the top left of each panel, in-line with the galaxy name)}, which also determines the colour of the individual points (with a colour bar on the right).
\fig{RHIMSpanels} is rich in physical information, and we dissect this in the following analysis. 

We first focus on the top row, which contains VERTICO galaxies considered to be \HI-normal based on their def-\HI\ ($-0.3<\text{def-\HI\ [dex]} <0.3$).
Three galaxies {are} consistent with the HERACLES control sample, while NGC 4567 appears to be a slight outlier, particularly at {lower} $\SigMstar$.
NGC 4567 is interacting with NGC 4568 (row 2, column 2) and the galaxies cannot be separated in velocity space  \cp{chung09}, causing an overlap of the \rHI\ relation of each galaxy and meaning that it has experienced some environmental mechanisms.

The \rHI\ relations of individual galaxies show two clear trends as a function of increasing def-\HI\ that occur distinctly at {lower} and {higher} $\SigMstar$. 
First, the value of $\lgSigMstarTR$ ({measuring $\RHI$}, vertical dashed lines), on average, increases with increasing def-\HI.
In other words, the \HI\ is  stripped (at least to our sensitivity limit) {to smaller radii} within the disc as def-\HI\ increases. 
Second, inside $\RHI$, the individual \rHI\ relations increasingly fall below the median and then the 16$^\mathrm{th}$ (lower) percentile of the HERACLES control sample at fixed $\SigMstar$.
That is, there is {an} increasing {reduction} of the typical $\SigHI$ within the $\RHI$ of galaxies as def-\HI\ increases.
This effect is largest in the {highest} def-\HI\ systems, but several moderate def-\HI\ galaxies also {begin to show evidence for this reduction} (NGC 4321, NGC 4501, NGC 4535). 
These trends show that the impact of {environmental processes} on the \HI\ is two-fold: not only is the \HI\ stripped from the outskirts, but the typical {surface} density of the remaining gas {is also reduced}.
{Additionally, we observe no evidence for an environment-driven enhancement of $\SigHI$, at least at the 1.2-kpc spatial scale that we trace.}

To quantify these effects, and because this is a complex parameter space, we use two different strategies for computing and comparing  $\lgSigHI$ histograms of VERTICO galaxies to the control sample. 
In all comparisons, we split the VERTICO galaxies into two bins of {global} def-\HI\footnote{{As mentioned in \S\ref{subsec:sample}, our def-\HI\ bins are wide enough to fully encapsulate any small, second-order correlation of $\SigHI$ and def-\HI\ due to the variation of $\SigHI$ profiles in galaxies, guaranteeing that our analysis is tracing environmentally-related processes.}}, def-\HI\ $=[0.3,0.7)\, \mathrm{dex}$ and def-\HI\ $\geq0.7 \, \mathrm{dex}$, which we refer to these as the `moderate' and `large' def-\HI\ samples, respectively.
{These thresholds were defined as approximately 1 and 2 times the scatter in the def-\HI\ of isolated galaxies ($\sigma$=0.3\,dex), with a slight adjustment to ensure an equal number of galaxies in each sample.}
Histograms are computed by weighting each galaxy equally {(see} \S\ref{sec:RMScomb}), and we also computed weighted medians for each def-\HI\ sample and the control sample, which are presented in Table \ref{tab:weighted_stats}.


\begin{figure}[t!]
    \centering
    \includegraphics[width=0.9\textwidth]{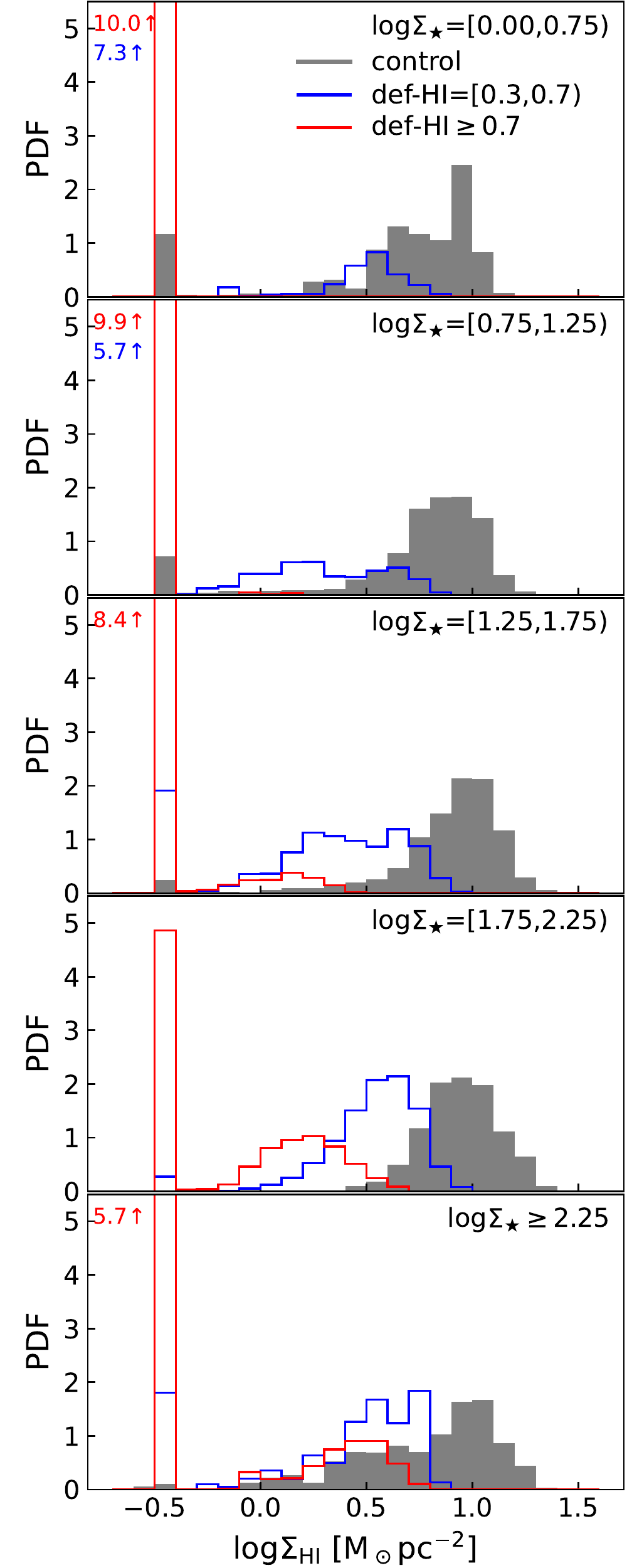}
    \caption{$\lgSigHI$ histograms in bins of $\lgSigMstar$.
    Each panel shows density-normalised histograms of $\lgSigHI$ for spaxels in the $\SigMstar$ range quoted in the top right corner. 
    HERACLES control sample galaxies are shown with a grey, filled histogram, VERTICO moderate def-\HI\ galaxies  with a blue,  open histogram, and large def-\HI\ galaxies with a red, open histogram. 
    Non-detected spaxels are included in the bin centered at $\lgSigHIMsun=-0.5$, and if the value of this bin exceeds the y-axis limit then its value is quoted in the top-left of the panel in the same colour as its histogram.
    At fixed $\SigMstar$, galaxies with larger def-\HI\ have a higher fraction of non-detected spaxels and their detected spaxels have lower $\SigHI$.
    The difference is larger at {lower} $\SigMstar$.}
    \label{fig:HI_Mstarbins}
\end{figure}

In our first comparison, shown in \fig{HI_Mstarbins}, we divide the \rHI\ relation into 0.5\,dex bins of $\lgSigMstar$ and compute the {histogram of} $\lgSigHI$ in each bin.
The control sample is shown as a grey, solid histogram, the moderate def-\HI\ population as a blue, open histogram, and the large def-\HI\ as a red, open histogram.
All \HI\ non-detected spaxels are included in the bin located at $\lgSigHIMsun=-0.5$, and if the value of this bin exceeds the figure limits, its value is written in the top left. 

\fig{HI_Mstarbins}  again highlights the trends seen in \fig{RHIMSpanels}.
First, both def-\HI\ samples show a larger fraction of non-detected \HI\ spaxels than the control sample, and the fraction of non-detected spaxels is higher in the larger def-\HI\ galaxies.
Namely, {the radius at which  \HI\ is {below the detection limit} gets closer to the galaxy centre} as def-\HI\ increases.
This increased fraction of non-detections is also evident in Table \ref{tab:weighted_stats}, where the weighted median for detections only is often different to the value for all points in both def-\HI\ samples.
Second, at all $\SigMstar$, the typical detected $\SigHI$ is also {lower} as def-\HI\ increases.
Table \ref{tab:weighted_stats} shows that the moderate def-\HI\ sample has a weighted median for detected $\SigHI$ spaxels that is $0.34-0.83$\,dex {lower} than the control sample, while it is $0.53 - 0.85$\,dex {lower} for the large def-\HI\ sample, {and these differences are greater than the estimated uncertainties.}
{Last, neither def-\HI\ sample shows {higher} $\SigHI$ than the control sample, highlighting that we observe no environment-driven enhancement of $\SigHI$}. 

{We note that there is some overlap between the 1\,$\sigma$ scatter in the $\SigHI$ distributions.}
{For example, in the highest $\SigMstar$ bin the 16$^\mathrm{th}$ percentile of the control sample (0.42) is smaller than the 84$^\mathrm{th}$ percentiles of both the moderate (0.73) and large (0.49) def-\HI\ samples.}
{As we do not expect environmental mechanisms to operate across the whole gas disc with the same magnitude (e.g., the centre or the trailing edge will be less affected), some overlap between the distributions is expected.}
{Nevertheless, in all $\SigMstar$ bins we find that the two distributions cannot be drawn from the same sample\footnote{{As we are using weighted data, we used a weighted K-S test, and our implementation can be found \href{https://github.com/awattsup/astro-functions/blob/d340fb20301294671c5eaa0857e613c59945f239/astro_functions.py\#L430}{here}.}} ($p<10^{-4}$), confirming that we are witnessing the effect of environmental processes already in the moderate def-\HI\ regime.}


{Considering the K-S test results,} \fig{HI_Mstarbins} also contains interesting insights into how {environmental processes} change the {distribution of $\SigHI$ in} the remaining gas. 
While the {bins with} $\lgSigMstarMsun<1.25$ are dominated by non-detections, in the {higher} $\lgSigMstar$ bins, the $\lgSigHI$ histogram of the moderate def-\HI\ galaxies (blue) is similar to the control sample.
Namely, there is a peak at {higher} $\SigHI$ and a tail toward {lower} {surface} densities. 
The main difference is a decrease in the average $\SigHI$, while the increase in non-detections comes from the low {surface} density tail falling below the detection limit. 
The large def-\HI\ galaxies show a larger decrease in $\SigHI$ values, but also, the shape of the $\lgSigHI$ histogram becomes more gaussian rather than skewed. 
This change in shape suggests that in addition to the decrease in the typical surface density, the densest \HI\ regions {are no longer present}.

\begin{table*}[t!]
	\centering
	\caption{Weighted median $\lgSigHI$ (\fig{HI_intr}) and $\lgSigMG$ (\fig{MG_intr}) within $\RHI$ for the   moderate and large def-\HI\ samples, and their matched control samples.
		As the control sample histogram is constructed by selecting spaxels with $\lgSigMstar\geq\lgSigMstarTR$ in each galaxy, the two def-\HI\ samples have unique control samples, and thus {unique} weighted control sample medians.  
	The first column gives the def-\HI\ sample, the second and third columns the control sample and def-\HI\ sample medians using all spaxels, and the remaining columns use detected spaxels only.
	Uncertainties are calculated using jackknife re-sampling and rejecting an entire galaxy in each iteration.}
	\label{tab:intr_stats}
	\begin{tabular}{lccccc} 
		\hline \hline
		Sample  & Control median $\lgSigHI$ & Sample median $\lgSigHI$  & Control median detected  $\lgSigHI$ & Sample median detected  $\lgSigHI$ \\
		 \cmidrule(lr){2-3}\cmidrule(lr){4-5}
		\hline
		def-\HI$=[0.3,0.7)$ & $0.95 \pm 0.02$  & $0.50 \pm 0.07$  & $0.95 \pm 0.02$ & $0.52 \pm 0.07$  \\
		def-\HI$\geq0.7$ & $0.94 \pm 0.02$  & $0.19 \pm 0.14$  & $0.94 \pm 0.02$ & $0.28 \pm 0.06$\\
		\hline \hline
		Sample & Control median $\lgSigMG$ & Sample median $\lgSigMG$  & Control median detected $\lgSigMG$   & Sample median detected $\lgSigMG$  \\
		\cmidrule(lr){2-3}\cmidrule(lr){4-5}
		\hline
		def-\HI$=[0.3,0.7)$ & $0.88 \pm 0.03$ & $0.77 \pm 0.08$  & $0.95 \pm 0.02$ & $0.80 \pm 0.07$  \\
		def-\HI$\geq0.7$ & $1.15 \pm 0.04$ & $0.70 \pm 0.09$  & $1.16 \pm 0.04$ &  $0.71 \pm 0.09$\\
		\hline \hline
    \end{tabular}
\end{table*}
\begin{figure}[t!]
    \centering
    \includegraphics[width=0.9\textwidth]{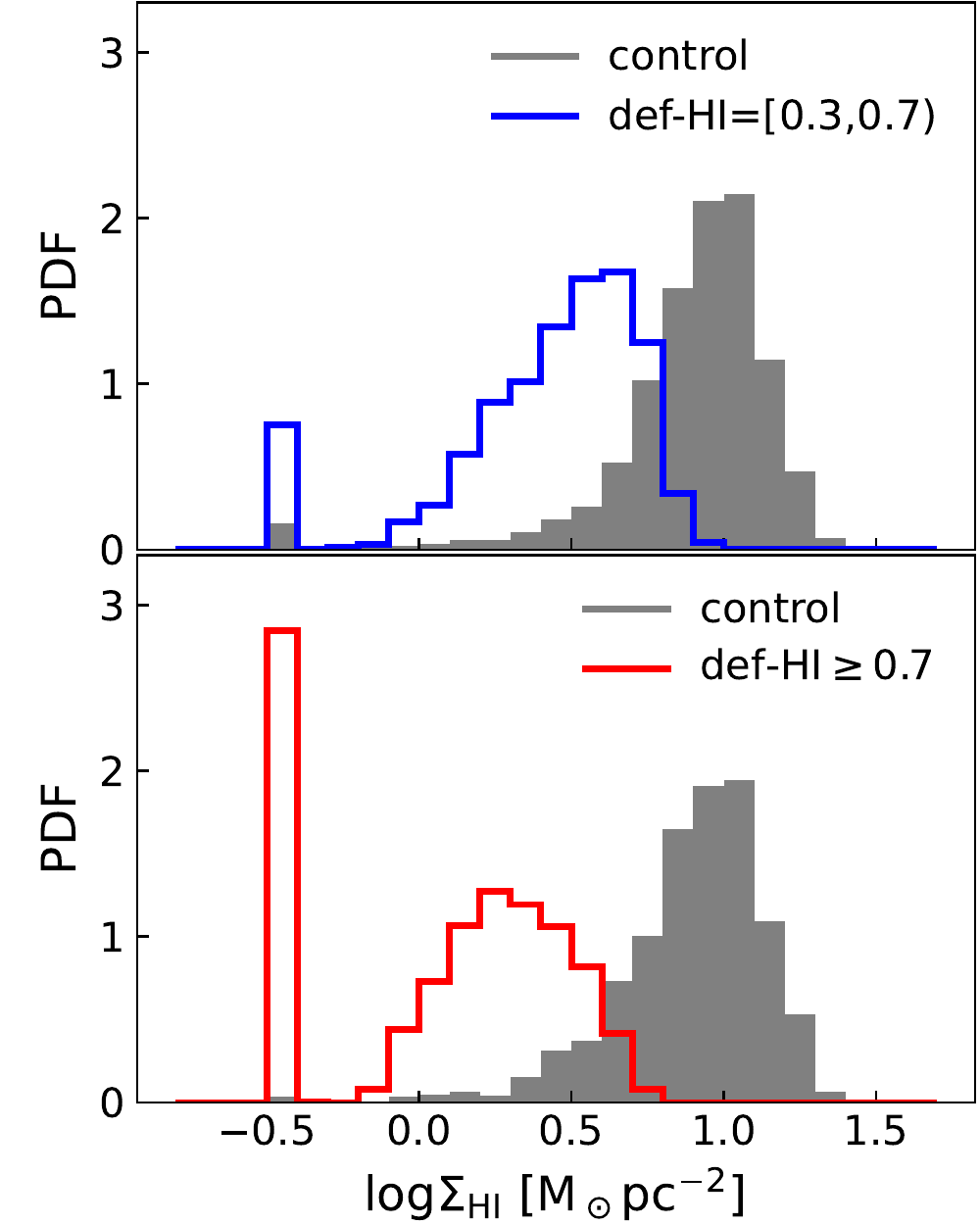}
    \caption{{Histograms of} $\lgSigHI$ within the $\RHI$ of galaxies. 
    We compare density-normalised histograms of $\lgSigHI$ for spaxels with $\lgSigMstar\geq \lgSigMstarTR$ for galaxies with moderate def-\HI\ (top panel; blue, open histogram) and large def-\HI\ (bottom panel; red, open histogram) to their matched HERACLES control samples (both panels, grey filled histogram).
    As def-\HI\ increases, the fraction of non-detected spaxels within $\RHI$ increases, and the typical detected $\SigHI$ is {lower}. 
    In galaxies with the largest def-\HI\, the {histogram of} detected $\lgSigHI$ is gaussian, rather than skewed, indicating that {the densest \HI\ is no longer present}. 
    {Neither def-\HI\ sample shows evidence for elevated $\SigHI$ compared to the control sample.}
    }
    \label{fig:HI_intr}
\end{figure}

In our second comparison, we {consider} only \HI\ within the $\RHI$ of each galaxy.
For both def-\HI\ samples, we used all spaxels with $\lgSigMstar \geq \lgSigMstarTR$ to compute equally-weighted density-normalised $\lgSigHI$ histograms.
Achieving a fair comparison with the control sample required an extra step, however, as the $\lgSigMstarTR$ of each galaxy is different. 
For each VERTICO galaxy, we computed a unique, equally-weighted control sample histogram using all spaxels from each control galaxy with $\lgSigMstar \geq \lgSigMstarTR$. 
The control histograms for each galaxy in a given def-\HI\ sample were then stacked and re-normalised using the same procedure\footnote{{This was done by introducing an additional factor of $1/D$ to the weight for each spaxel in each unique control sample histogram, where $D$ is the number of galaxies in the def-\HI\ sample.}}.
This dependence on $\lgSigMstarTR$  results in slightly different control samples for each def-\HI\ sample\footnote{We tested a different approach of combining all spaxels from the control galaxies and found the same results for the \HI\ using this method. This is because $\SigHI$ is essentially flat with $\SigMstar$, so there is little dependence on the $\SigMstar$ range covered. The same is not true for the molecular gas, so for consistency, we use the same approach for both gas phases.} and is visible when comparing the grey histograms in the two panels of \fig{HI_intr}. 
{Weighted medians for these histograms are listed in Table \ref{tab:intr_stats}.}

In \fig{HI_intr}, we compare histograms of  $\lgSigHI$ for spaxels within  $\RHI$ between the control sample (grey, solid), the moderate def-\HI\ sample (top panel, blue) and large def-\HI\ sample (bottom panel, red). 
Here, we can see the impact of {environmental processes} on the remaining \HI\ gas that has not been removed.
The moderate def-\HI\ galaxies have median detected $\lgSigHIMsun=0.52\pm0.07$, while their matched control sample has $\lgSigHIMsun=0.95\pm0.02$, indicating that the typical  $\SigHI$ is 0.43\,dex {lower} (Table \ref{tab:intr_stats}). 
The difference is larger in the large def-\HI\ galaxies, which have a median $\lgSigHIMsun=0.28\pm0.06$ that is 0.66\,dex lower than the median of their matched control sample, $\lgSigHIMsun=0.94\pm0.02$.
We also see the same changes in the shape of the $\SigHI$ histogram described above in \fig{HI_Mstarbins}, and we can rule out the  control and def-\HI\ samples being drawn from the same parent distribution ($p<10^{-4}$). 
Thus, when considering the remaining gas inside  $\RHI$,  {environmental processes} typically lower the average $\SigHI$ of moderate def-\HI\ galaxies, while large def-\HI\ galaxies can no longer support the densest \HI\ gas.

Our results so far have provided a unique view of how {environmental mechanisms} impact the resolved \HI\ content of galaxies, and agree with results from previous studies \cp[e.g.][]{vollmer01,tonnesen09,chung09,yoon17,cortese21,boselli22,lee22}.
However, how these mechanisms impact the molecular gas in galaxies on these spatial scales {has so far remained} unquantified, {and is discussed next}.

\subsection{Molecular hydrogen} \label{subsec:RMGMS}
\begin{figure*}[t!]
    \centering
    \includegraphics[width=\textwidth]{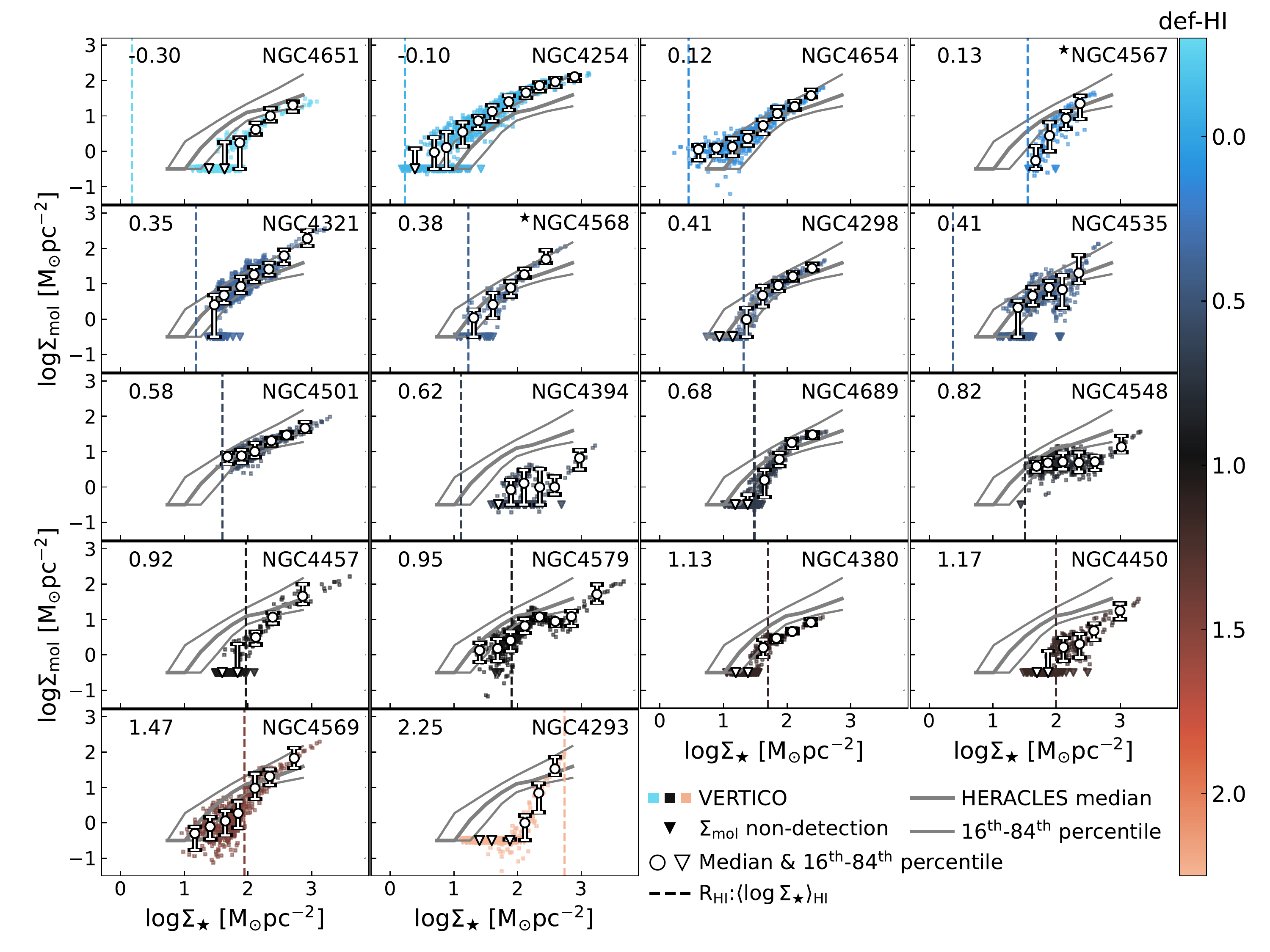}
    \caption{
    Individual \rMG\ relations for VERTICO galaxies, presented in the same way as \fig{RHIMSpanels}.
    {As def-\HI\ increases, the clearest deviation of $\SigMG$ from the HERACLES control sample occurs at  {lower} $\SigMstar$, {typically} close to {or outside of} $\RHI$ ($\lgSigMstarTR$).}
    Only some of the galaxies with the largest def-\HI\ have {{reduced} $\SigMG$ at {higher} $\SigMstar$ and within their $\RHI$} {(e.g., NGC 4579 NGC 4380, NGC 4450)}.
    }
    \label{fig:RMGMSpanels}
\end{figure*}
In \fig{RMGMSpanels}, we present the individual \rMG\ relations of VERTICO galaxies in the same way as the \rHI\ relations.
Focusing first on the top row of \HI-normal galaxies, aside from NGC 4567 (interacting, as discussed in \S\ref{subsec:RHIMS}), NGC 4651 seems to be molecule-poor for an \HI-normal galaxy. 
The kinematic position angles of the \HI\ inside and outside the optical disc of NGC 4651 are different \cp{chung09}, and it has stellar and \HI\ tidal features \cp{chung09,martinez-delgado10,morales18}, suggesting that much of its \HI\ disc could have been recently accreted.
Thus, the molecular gas content likely does not reflect that of an \HI-normal galaxy.
NGC 4254 appears slightly molecule-rich at fixed $\SigMstar$.
There is evidence that it recently underwent a fly-by interaction \cp{vollmer05,haynes07,boselli18a}, and it has a molecular bar and asymmetric spiral arms containing dense molecular gas \cp{sofue03}.
{These differences between the control sample and NGC 4651 and NGC 4254 highlight how environmental mechanisms can influence galaxies before they become significantly \HI-poor, and the importance of selecting a control sample from field galaxies.}

\begin{figure}[t!]
    \centering
    \includegraphics[width=0.9\textwidth]{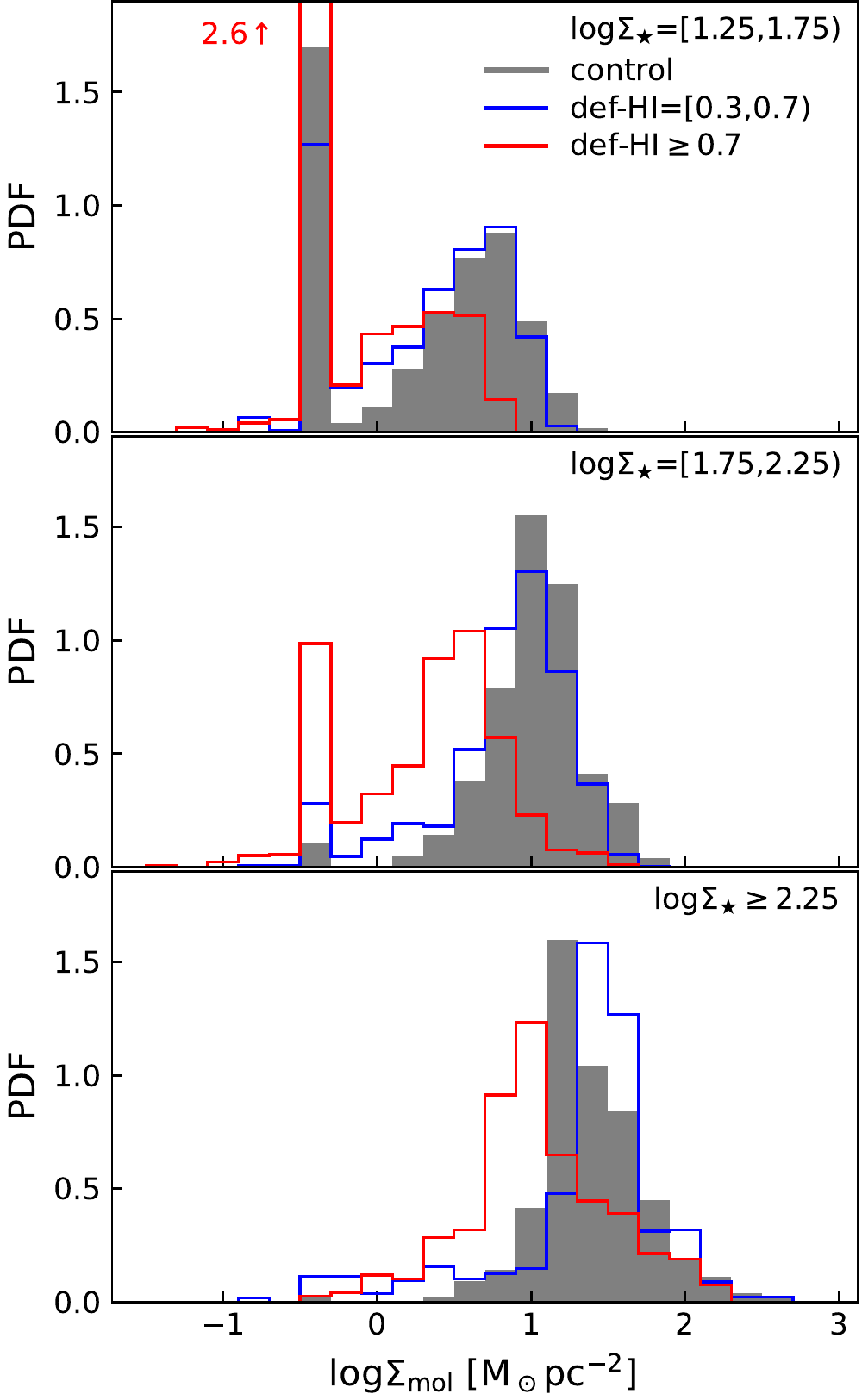}
    \caption{
    {Histograms of} $\lgSigMG$ in bins of $\SigMstar$, presented in the same way as \fig{HI_Mstarbins}.
    At fixed $\SigMstar$, moderate def-\HI\ galaxies have a similar {range} of detected $\SigMG$ to HERACLES galaxies, with a slight extension to {lower} $\SigMG$, while large def-\HI\ galaxies show a {lower} typical $\SigMG$. 
    }
    \label{fig:MG_Mstarbins}
\end{figure}

Before dissecting \fig{RMGMSpanels}, it is useful to point out some interesting features.
First, the median \rMG\ relation of four galaxies (NGC 4298, NGC 4689, NGC 4457, NGC 4450) transition from detections to non-detections coincidently with  $\RHI$, $\lgSigMstarTR$, indicating that these galaxies show {co-spatial} \HI\ and molecular gas truncation. 
Second, several galaxies show changes in the slope of their \rMG\ relation.
Some of these changes are close to $\RHI$ and are likely the signature of the environment (e.g. NGC 4298, NGC 4689).
Other galaxies show changes in their more central regions, such as the superposition of two slopes in NGC 4535, a flattening and then upturn in NGC 4548, or the zig-zag structure of NGC 4579. 
These are not solely due to changes in the $\SigMG$ content but also reflect changes in the central stellar structure of galaxies.
NGC 4535 and NGC 4548 are barred with central bulges, causing a steeper $\SigMstar - \SigMG$ relation in their centres and bar regions and a flatter relation in the disc.
NGC 4579 has a bar and a ring, causing an alternation between elevated $\SigMG$ and $\SigMstar$ in the centre,  lowered $\SigMG$ in a {higher} $\SigMstar$ region between the centre and the ring, and {higher} $\SigMG$ in a lower $\SigMstar$ region in the ring and disc.
Beyond having interesting {effects on} the shape of the \rMG\ relation, this suggests that we must use caution when comparing $\SigMG$ properties at fixed $\SigMstar$, particularly in the ranges where stellar structures could be important.

Comparing the \rMG\ relations of VERTICO galaxies to the control sample, there is less difference as a function of def-\HI\ than was visible for the \rHI\ relation. 
In the second and third rows, namely at moderate def-\HI, the \rMG\ relations of individual galaxies typically agree well with the median of the control sample, though there are some small signatures of deviation near $\RHI$ in some galaxies.
Instead, it is primarily the largest def-\HI\ systems (the bottom two rows) where the individual \rMG\ relations begin to deviate below the control sample median and occasionally below the 16$^\mathrm{th}$ percentile,  at fixed $\SigMstar$. 
{These differences are largest at {lower} $\SigMG$ and closer to $\RHI$, though there is galaxy-galaxy variation and at high  $\SigMstar$ some galaxies are consistent with the control sample (e.g. NGC 4457, NGC 4569), while others show {reduced} $\SigMG$ (e.g. NGC 4380, NGC 4450).}

These results imply that the $\SigMG$ at {lower} $\SigMstar$ ({outer parts of the disc}) is more strongly impacted than at {higher} $\SigMstar$ ({the inner regions}). 
Further, as $\SigMG$ is typically {lower} at {lower} $\SigMstar$, the {environmental processes} preferentially impact the lower {surface} density molecular gas within galaxies. 
Interestingly, this can begin even at moderate def-\HI\ (e.g. NGC 4298 and its $\SigMG$ slope-change and $\SigMG$ truncation) and effectively changes the slope of the individual \rMG\ relations, making them steeper.
{These results agree with previous studies, which found steeper radial profiles of $\SigMG$ in galaxies with more disturbed \HI\ morphology and larger def-\HI\ \cp{fumagalli09,zabel22,villanueva22}}.
{Additionally, like $\SigHI$ we observe no environment-driven elevation of $\SigMG$ at the spatial scale that we are sensitive to.}

In \fig{MG_Mstarbins} we compare the $\lgSigMG$ histograms of VERTICO galaxies to the control sample in 0.5\,dex $\lgSigMstar$ bins.
We do not expect stellar structures, as mentioned above, to cause any biases in these histograms as they are primarily contained in the $\lgSigMstarMsun\geq2.25$ bin.  
These histograms reflect the smaller offsets between the individual VERTICO \rMG\ relations and the control sample. 

{Moderate def-\HI\ galaxies (blue) show little differences from the control sample, except in the lowest $\SigMstar$ bin that has a significant contribution from non-detections.}
{In this bin, the relatively larger number of non-detected spaxels in the control sample is due to the shallower detection limit of HERACLES.}
{Overall}, the main difference {between the control sample and moderate def-\HI\ sample} is the presence of a slightly more extended tail toward {lower} $\SigMG$. 
This tail imprints on the median detected $\SigMG$ being up to $\sim$0.18 dex {lower} than the control sample, but the {distributions peak in similar locations and in the two higher $\SigMstar$ bins the medians are consistent} within their estimated uncertainties (Table \ref{tab:weighted_stats}).}
{The presence of this tail, however, is sufficient to rule out the control sample and moderate def-\HI\ sample being drawn from the same distribution in all $\SigMstar$ bins ($p<4\times10^{-4}$).}

In the large def-\HI\ sample (red), the impact of {environmental processes} is stronger.
Non-detections dominate the smallest $\SigMstar$ bin, {and despite the VERTICO data being deeper, the relative number of non-detected spaxels is larger than the control sample in this bin.}
Considering the detected spaxels, the weighted median $\lgSigMG$ is  $0.34-0.59$\,dex {lower} than the control sample across all $\lgSigMstar$ bins, {all of which are significant considering their uncertainties} (Table \ref{tab:weighted_stats}), {and we can exclude that the distributions are drawn from the same parent sample ($p<10^{-4}$)}.

These results support the interpretation that it is primarily the lower {surface} density molecular gas at {lower} $\SigMstar$ that is affected by the environment, although {as galaxies become significantly \HI-deficient ($\gtrsim0.7$)}, the impact is larger and can also affect the gas at {higher} $\SigMstar$.
{Further, in the histograms of \fig{MG_Mstarbins} we see no evidence for any {environment-driven} elevation  of $\SigMG$, {although} in the largest $\SigMstar$ bin the moderate def-\HI\ sample appears to show slightly {higher} $\SigMG$}.
{However, the medians are consistent within the uncertainties, and this enhancement disappears if we instead consider spaxels with $\lgSigMstarMsun\geq2$ or $\lgSigMstarMsun=[2,2.5)$, meaning that it is due to the slightly different $\SigMstar$ coverage of galaxies in this bin.}

\begin{figure}[t]
    \centering
    \includegraphics[width=0.9\textwidth]{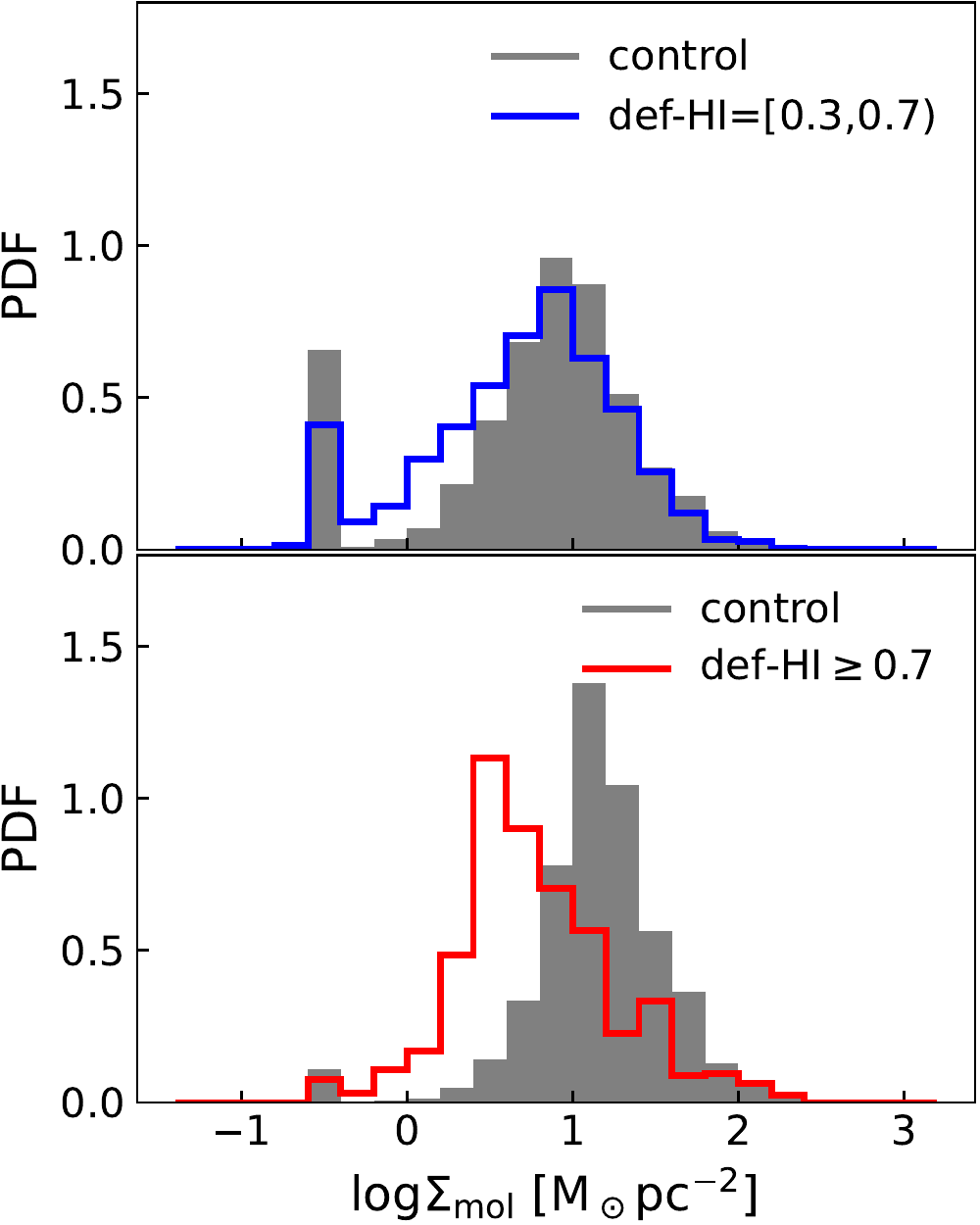}
    \caption{
    {Histograms of} $\lgSigMG$ within the $\RHI$ of galaxies, presented in the same way as \fig{HI_intr}.
    Inside the $\RHI$ of moderate def-\HI\ galaxies,  {environmental processes} affect primarily {the molecular gas with low surface density}, while in large def-\HI\ galaxies {even} the typical $\SigMG$ is {reduced}.
    }
    \label{fig:MG_intr}
\end{figure}

In \fig{MG_intr}, we compare the histograms of $\lgSigMG$ within $\RHI$ for moderate def-\HI\ galaxies (top, blue) and large def-\HI\ galaxies (bottom, red) to their respective control sample $\lgSigMG$ histograms{\footnote{{As the molecular gas observations of some galaxies do not extend to $\RHI$ (e.g., NGC 4321, NGC 4535), we picked the minimum $\SigMstar$ of each VERTICO galaxy (used to select control sample spaxels) to be the largest out of $\lgSigMstarTR$, $\lgSigMstarMsun=1.25$ (see \S\ref{sec:RMScomb}), and the minimum $\lgSigMstar$ for the galaxy. In this way, only the observed $\SigMstar$ range is used to compute the weighted control sample.}}} (grey, solid). 
The $\SigMG$ histogram of moderate def-\HI\ galaxies is similar to the control sample, and the median detected $\SigMG$ is $0.15$\,dex {lower} (a factor of $\sim$1.4, Table \ref{tab:intr_stats}), {which is significant considering the uncertainty}. 
Much like \fig{MG_Mstarbins}, the main difference is the tail toward lower $\SigMG$ values, and there is also a {reduction of spaxels with} surface densities near the peak of the histogram.
{The control sample has slightly more non-detected spaxels due to the shallower detection limit, and we verified that the result remains if consider only spaxels with $\lgSigMstarMsun\geq1.5$.}
{We also note that the tail is primarily driven by NGC 4394, which sits below the control sample \rMG\ relation at all $\SigMstar$.}
{We re-computed the $\lgSigMG$ histogram with NGC 4394 excluded and found that a smaller tail toward lower {surface} density $\SigMG$ still remains.}

The $\lgSigMG$ histogram of the large def-\HI\ galaxies shows a different behaviour.
Unlike the moderate def-\HI\ galaxies, there is no extended tail of {lower} $\SigMG$, but instead, it has a similar shape to the control sample, only shifted to a {lower} average $\SigMG$. 
The weighted median {detected spaxel} is $0.45$\,dex {lower} {(a factor of 2.8 {or 64 per cent})} than the control sample {and significant considering the uncertainties} (Table \ref{tab:intr_stats}), reflecting a larger change, which would be expected in larger def-\HI\ galaxies. 
Thus, in the largest def-\HI\ galaxies, {environmental processes act} to lower the typical $\SigMG$ within  $\RHI$.
These results support the interpretation that at moderate def-\HI , the lower {surface} density molecular gas within  $\RHI$ is impacted, while at large def-\HI\ the typical $\SigMG$ is lowered. 

To summarise, \emph{we have presented clear evidence that {environmental processes} impact the molecular gas content of galaxies on 1.2-kpc physical scales, {even within the \HI\ truncation radius.}}
The effect is primarily on the lower {surface} density molecular gas and can begin even at moderate def-\HI , while in the largest def-\HI\ systems, the denser molecular gas shows a reduction in the average molecular gas {surface} density.

\section{Physical drivers of $\SigMG$ {{reduction}}} \label{sec:MGHIr}

{The observed {{reduction}} of $\SigMG$ at fixed $\SigMstar$ suggests that {cluster} {environmental processes} are impacting the molecular gas in VERTICO galaxies.}
{This reduction could be due to direct stripping of {the} molecular gas, similar to the \HI\ \cp[e.g.,][]{moretti18,cramer20}.}
Alternately, it could result from {inefficient molecule formation caused by} a change in the ISM physical conditions due to the {stripping} of the \HI\ \cp[e.g.,][]{blitz06,krumholz09,fumagalli09}.
{In this case}, the suppressed molecular gas content is due to consumption by star formation {and the lack of cold ISM replenishment ({akin to} strangulation/starvation)}.
{However, for this last scenario to work requires the assumption that the ISM in a galaxy is able to respond to any changes on a timescale equal to, or shorter than, the stripping timescale of the \HI\ \cp[$\lesssim200\,$Myr;][]{vollmer01, roediger07,tonnesen09,choi22}}.
{Indeed, this might be expected due to the short lifetimes of GMCs \cp[$\lesssim30$\,Myr,][]{chevance20a}.}
{In this case, VERTICO galaxies should follow {the} scaling relations used to interpret the resolved {molecular-to-atomic gas mass ratio} ($\SigMGHI$) in normal galaxies, {which is} sensitive to the physical conditions within the ISM.}
{Instead, we show {here} that these scaling relations break down when applied to VERTICO galaxies, and they cannot describe their $\SigMGHI$, suggesting that changes to the ISM physical conditions {alone} cannot explain the observed {reduction} of $\SigMG$.}

\begin{figure*}[t!]
    \centering
    \includegraphics[width=\textwidth]{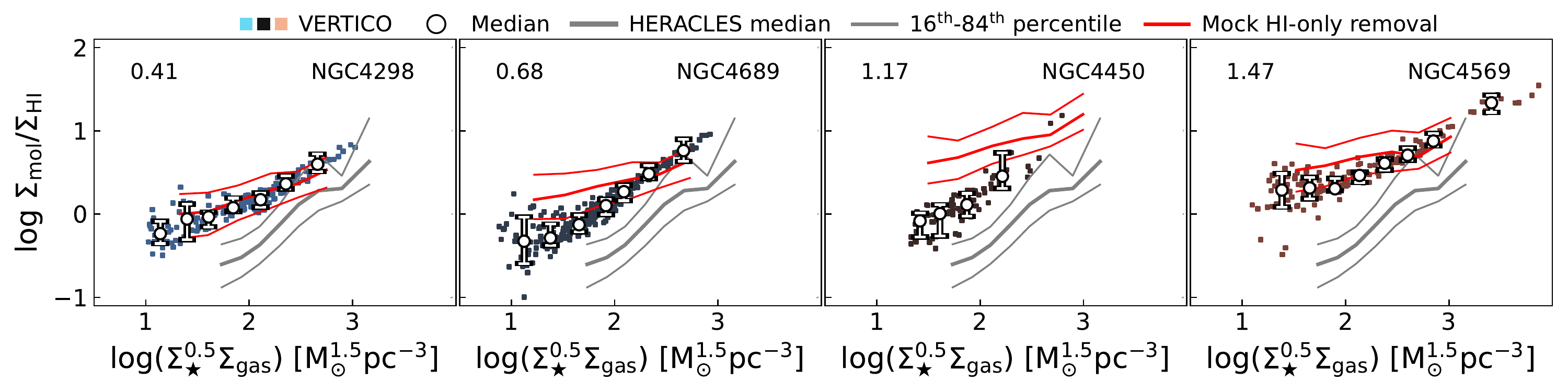}
    \caption{Ratio of molecular-to-atomic gas mass of individual spaxels as a function of the midplane pressure proxy $\SigMstar^{0.5}\SigGas$ for four VERTICO galaxies, {coloured and ordered by def-\HI}.
    Galaxies are presented in the same format as in \fig{RHIMSpanels} and \fig{RMGMSpanels}, with the left two belonging to the moderate def-\HI\ sample and the right two to the large def-\HI\ sample. 
    {The red lines show how the median and 16$^\mathrm{th}$ and 84$^\mathrm{th}$ percentiles of the HERACLES control sample would shift if it had experienced the same magnitude of \HI\ {reduction} at fixed $\SigMstar$ as each VERTICO galaxy.}
    At fixed midplane pressure, VERTICO galaxies have a larger $\SigMGHI$ than expected for the same pressure in a control sample galaxy. {However, this difference is not as large as the \HI-only removal scenario, particularly at {lower} pressures and larger def-\HI}.
    }
    \label{fig:MGHIr_MPP}
\end{figure*}

\begin{figure}[t!]
    \centering
    \includegraphics[width=0.9\textwidth]{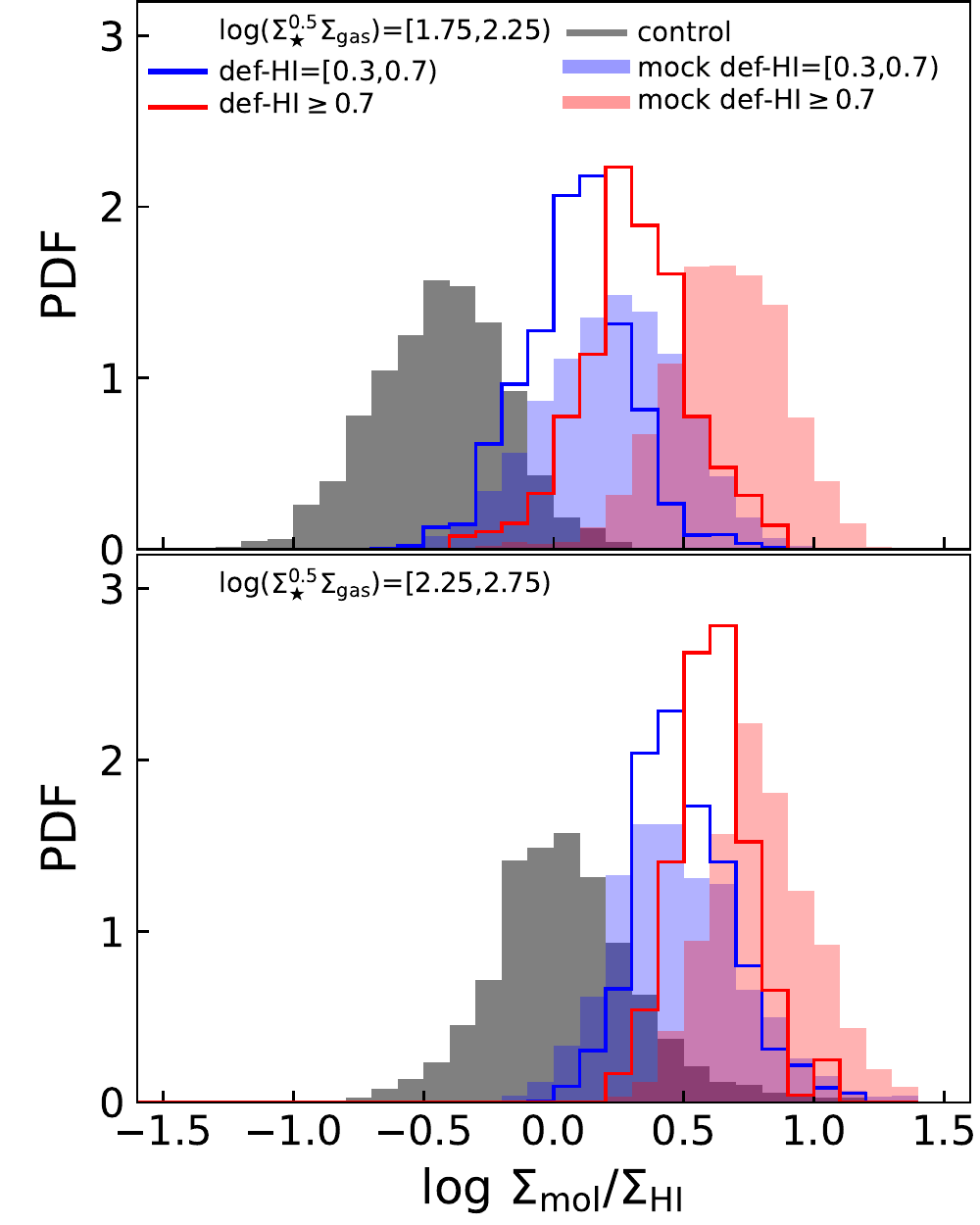}
    \caption{{Histograms of} $\lgSigMGHI$ in bins of midplane pressure, presented in the same way as \fig{HI_Mstarbins} and \fig{MG_Mstarbins}.
    {Additionally, solid blue and red histograms show how the $\lgSigMGHI$ histogram of the control sample would move if it had experienced \HI\ removal of equal magnitude as the moderate and large def-\HI\ samples, respectively.}
    Moderate and large def-\HI\ galaxies have systematically larger $\SigMGHI$ than control sample galaxies at fixed $\SigMstar^{0.5}\SigGas$.
    {Further, the \HI-only removal scenario over-estimates the observed $\SigMGHI$, except in moderate def-\HI\ galaxies in the largest $\SigMstar^{0.5}\SigGas$ bin.}
    }
    \label{fig:MGHIr_MPPbins}
\end{figure}

More detailed modelling of the physical state of the cold gas reservoir, and its relationship to $\SigMGHI$, is outside the scope of this work and will be investigated in future VERTICO papers.
{For an analysis of $\SigMGHI$ across the discs of VERTICO galaxies using radial profiles, we refer the reader to  \ct{villanueva22}, who found elevated ratios in galaxies with perturbed and truncated \HI\ morphologies.}
{Here,} we compare $\SigMGHI$ in individual spaxels to two parameterisations {of the} ISM physical conditions to try and understand what physics might be responsible for the reduction in molecular gas content of the VERTICO galaxies.
{To be consistent} with previous works, in this section, we correct our \HI\ masses for a 36 per cent contribution from Helium.
{Our analysis is restricted to detected spaxels only, and we also only consider spaxels with $\lgSigMstar\geq1.25$ so that we only study the ISM parameterisations where the control sample is dominated by detected spaxels.}
We also denote the total cold gas {surface} density as $\SigGas=\SigHI+\SigMG$.

{It is also useful to quantify how the removal of \HI\ alone manifests {in the scaling relations that we will investigate}.}
{To do this, we computed the offset between the median $\lgSigHI$ of each VERTICO galaxy and the median of the control sample at fixed $\SigMstar$.}
{We then subtracted this offset (as a function of $\SigMstar$) from the spaxels of each HERACLES galaxy to create mock `\HI-stripped' galaxies.}
{Using these mocks, we re-computed weighted histograms and statistics to estimate how the HERACLES control sample would shift if it had experienced \HI\ reduction equal to what is observed for each VERTICO galaxy.}

{{We use these mock galaxies} in the following analysis to show where deviations from the ISM scaling relations can be explained by \HI-only removal, and where we also require extra reduction of the molecular gas content.}
{Indeed, it is our ability to perform this kind of spatially resolved analysis that allows us to identify these small differences, which are on the order of $0.1-0.2$\,dex, that would likely be washed out in studies considering only integrated quantities such as global \HI\ and molecular gas deficiencies.}
{For conciseness, we only show individual scaling relations for a subset of VERTICO galaxies in this section.
Individual relations for all VERTICO galaxies, and the combined VERTICO and HERACLES relations are shown in \ref{sec:app1}.}

\begin{figure*}[t!]
    \centering
    \includegraphics[width=\textwidth]{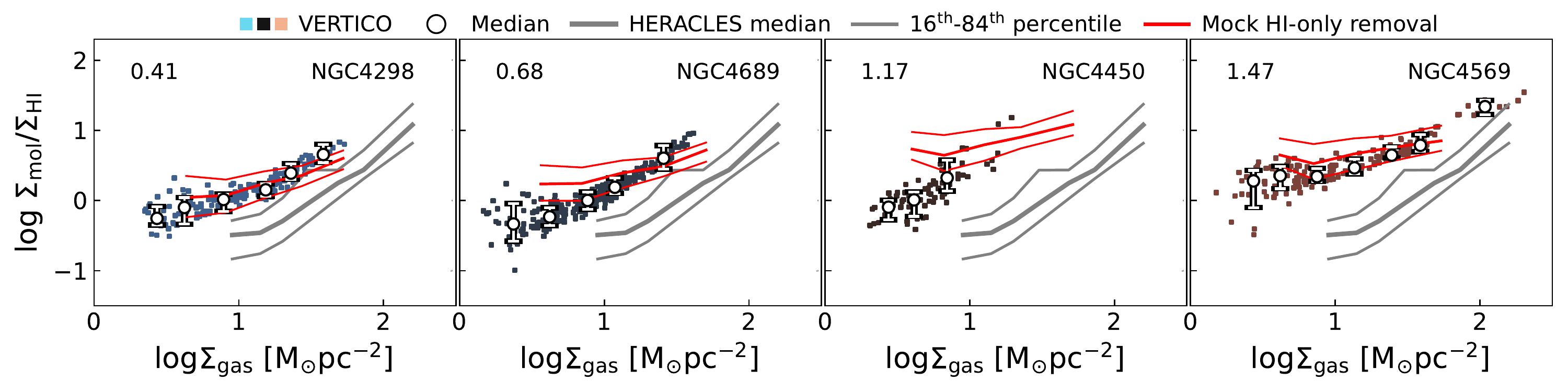}
    \caption{Ratio of molecular-to-atomic gas mass of individual spaxels as a function of $\SigGas$ for the same galaxies as \fig{MGHIr_MPP}.
    The $\SigMGHI$ of VERTICO galaxies is systematically larger than in control sample galaxies at fixed $\SigGas$, {but not as large as the \HI-only removal scenario shown with the red lines}.
    }
    \label{fig:MGHIr_CG}
\end{figure*}

\begin{figure}[t!]
    \centering
    \includegraphics[width=0.9\textwidth]{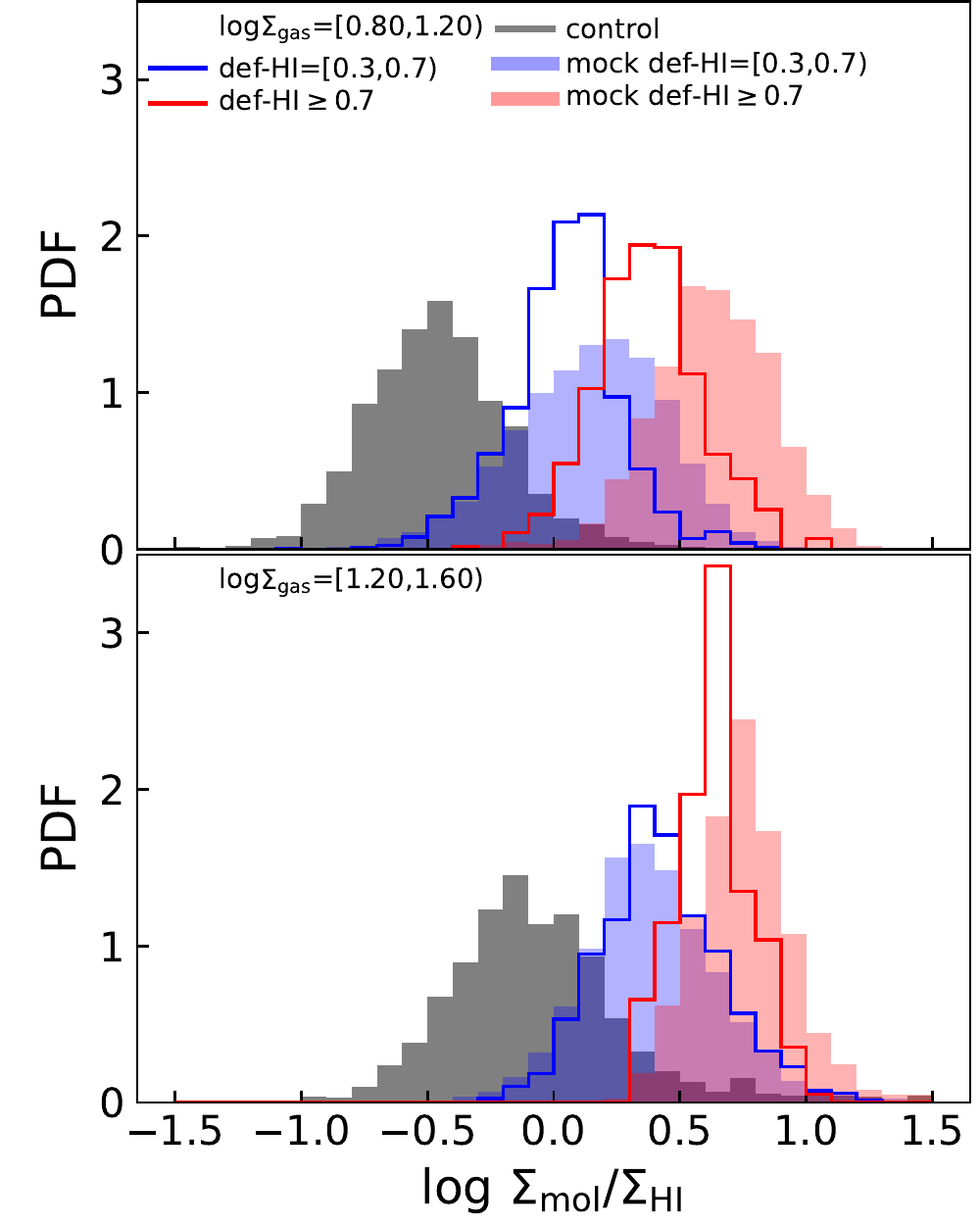}
    \caption{{Histograms of} $\lgSigMGHI$ in bins of total cold gas surface density, presented in the same way as \fig{MGHIr_MPPbins}.
    Moderate and large def-\HI\ galaxies have systematically larger $\SigMGHI$ than expected if molecular gas formation was set primarily by $\SigGas$.
    {The \HI-only removal scenario over-estimates $\SigMGHI$, except at large $\SigGas$ in moderate def-\HI\ galaxies.}
    }
    \label{fig:MGHIr_CGbins}
\end{figure}

First, we use a proxy for the hydrostatic pressure at the midplane of the disc, {which was predicted to determine the molecular gas fraction by \ct[][but see also \citealt{blitz06} and \citealt{leroy08}]{elmegreen93}}.
We adopt the prescription from \ct{blitz06}, which we estimate as $\SigMstar^{0.5}\SigGas$.
$\SigMstar$ and $\SigGas$ are the dominant terms in the \ct{blitz06} prescription, with the stellar scale height and gas velocity dispersion having minor contributions. 
If this prescription is an accurate predictor of physics determining $\SigMGHI$, then we might expect that stripping of the ISM (predominantly \HI) would reduce the pressure and cause $\SigMGHI$ to decrease correspondingly.

In \fig{MGHIr_MPP}, we show $\SigMGHI$ as a function of $\SigMstar^{0.5}\SigGas$ for two moderate def-\HI\ galaxies (left two panels) and two large def-\HI\ galaxies (right two panels), presented {similarly to} \fig{RHIMSpanels} and \fig{RMGMSpanels}.
{We also show, {with red lines}, how the HERACLES control sample median would shift if it had experienced \HI\ reduction equal to what is observed for the displayed VERTICO galaxy.}
All four VERTICO galaxies have larger $\SigMGHI$ than expected at fixed $\SigMstar^{0.5}\SigGas$ {showing that, despite a reduction in pressure due to the stripping of the \HI, the $\SigMGHI$ is \textit{larger}}. 
{Thus, {internal} hydrostatic pressure cannot explain the $\SigMGHI$ ratio in these VERTICO galaxies, suggesting that stripping of the gas (at least of the \HI) is more dominant than the hydrostatic pressure, {at least over the timescale traced by VERTICO galaxies}.}
{Further, there are regions within \fig{MGHIr_MPP} where the mock \HI-stripped control sample over-estimates the $\SigMGHI$ of VERTICO galaxies.}
{This over-estimation is {typically} larger at {lower} pressures (corresponding to larger radii) and in the larger def-\HI\ galaxies, which are the regions and galaxies where we observed molecular gas reduction in \S\ref{subsec:RMGMS}.}
{Thus, changes in the hydrostatic pressure plus  \HI\ removal alone cannot explain the observed molecular gas reduction in these galaxies.}

{In \fig{MGHIr_MPPbins}, we show that the story told by these four galaxies holds for our whole VERTICO sample, using normalised histograms of $\lgSigMGHI$ in two bins of $\log \SigMstar^{0.5}\SigGas$.}
{The blue and red {empty} histograms correspond to the moderate and large def-\HI\ samples, and the solid grey histogram shows the  $\SigMGHI$ histogram for the control sample.}
{The solid blue and red histograms show the mock \HI-stripped control samples, matched to the moderate and large def-\HI\ samples respectively.}
In both bins, the VERTICO galaxies have elevated $\SigMGHI$ compared to the HERACLES control sample; {but not so elevated that they match with the histograms of the mock \HI-stripped  control samples.}

{Quantitatively, in the lowest $\SigMstar^{0.5}\SigGas$ bin, an \HI-only removal scenario would result in {$\SigMGHI$ distributions with medians of $0.24\pm0.03$\,dex and $0.65\pm0.03$\,dex  for the moderate and large def-\HI\ populations, which over-estimate their observed values of $0.09\pm0.07$\,dex and $0.31\pm0.04$\,dex by 0.15\,dex and 0.34\,dex, respectively.}
{In the highest $\SigMstar^{0.5}\SigGas$ bin, the large def-\HI\ galaxies {similarly} show {0.2\,dex} smaller $\SigMGHI$ {elevation (median $0.61\pm0.04$\,dex) than expected from \HI\ removal alone ($0.81\pm0.04$\,dex)}.}
{A K-S test between the def-\HI\ samples and their respective mock control samples rules out that they are drawn from the same distribution ($p<10^{-4}$).}
{Conversely, the moderate def-\HI\ sample in this bin matches its mock control sample, which is reflected in the {expected and observed $\SigMGHI$ medians of $0.47\pm0.03$\,dex and $0.48\pm0.04$\,dex,} respectively.}
{This match between the mock control sample and the moderate def-\HI\ sample agrees with our results in \S\ref{subsec:RMGMS}, where no reduction in $\SigMG$ was observed in these galaxies.}
{Further, we can rule out that the elevated $\SigMGHI$ is due to ram pressure providing additional pressure support and {increasing the} conversion {rate} of \HI\ to molecular gas, as we found no evidence for elevated molecular gas content in any of our VERTICO galaxies (at least at our 1.2-kpc spatial scale).}
{All these points indicate that some reduction in $\SigMG$ is required to explain the $\SigMGHI$ of VERTICO galaxies, but this parameterisation {with} hydrostatic pressure alone does not provide a physical explanation.}

An alternate scenario based on ISM physics is that, due to the removal of \HI, there is insufficient fuel to form new molecular gas.
We test this {hypothesis} using the total gas {surface} density ($\SigGas$), suggested by \ct[][see also \citealt{fumagalli09}]{krumholz09} to be the primary regulator of $\SigMGHI$ with metallicity and the interstellar radiation field playing secondary roles. 
If the stripping of the ISM lowers the total {surface} density such that the formation of molecular gas is inefficient, we would expect VERTICO galaxies to lie along the same relation {between $\SigMGHI$ and $\SigGas$} as the control sample, {as claimed in \ct{fumagalli09}}. 
In \fig{MGHIr_CG}, we show $\SigMGHI$ as a function of $\SigGas$ in the same style as shown in \fig{MGHIr_MPP} for the pressure relation. 
The displayed moderate (left two panels) and large (right two panels) def-\HI\ galaxies do not lie along the same relation as the control sample. 
Further, the removal of \HI\ alone would over-estimate the observed $\SigMGHI$ at fixed $\SigGas$, {particularly in {lower} $\SigGas$ regions}.

This trend remains for the whole sample, and in \fig{MGHIr_CGbins}, we show histograms of  $\SigMGHI$ in two bins of $\lgSigGas$ in the same style as \fig{MGHIr_MPPbins}.
{In the {lower} $\SigGas$ bin, the expected {median} $\SigMGHI$ from \HI-only removal scenarios, $0.17\pm0.04$\,dex  and $0.63\pm0.04$\,dex,  over-estimate the observed values of  $0.05\pm0.07$\,dex $0.47\pm0.03$\,dex for the moderate and large def-\HI\ samples by 0.12\,dex and 0.16\,dex respectively.}
{In the {higher} $\SigGas$ bin, the elevation in $\SigMGHI$ of the large def-\HI\ sample is smaller by {0.11\,dex, with an expected median $0.74\pm0.03$\,dex compared to $0.63\pm0.04$\,dex observed, while the moderate def-\HI\ galaxies agree more closely with their mock control sample, having $0.38\pm0.03$\,dex expected and $0.41\pm0.05$\,dex observed.}}
{We also can exclude that both def-\HI\ samples are drawn from the same parent distribution as their respective mock control samples ($p<10^{-4}$).}
Thus, in VERTICO galaxies it is not $\SigGas$ that sets $\SigMGHI$, and explaining the observed $\SigMGHI$ requires a reduction in their molecular gas content that is not related to insufficient fuel or an inefficient conversion of \HI\ into molecular gas.

We must also consider if there are any systematic differences between how molecular gas is traced in VERTICO and HERACLES. 
As the data have been homogenised and both use the same constant $R_{21}$ and $\alpha_\mathrm{CO}$ to convert \CCOO\ luminosity to molecular gas mass, this scenario is unlikely.
Although $R_{21}$ and $\alpha_\mathrm{CO}$ have their own uncertainties, this homogenisation means that our results are actually re-scaled differences between the observed \CCOO\ emission in our samples. 
Thus, to explain the difference between the typical $\SigMG$ inside the $\RHI$ of large def-\HI\ galaxies would require {VERTICO galaxies to preferentially have} \CCOO-dark gas such that the observed \CCOO\ luminosity under-estimates the actual amount of molecular gas present by a factor of 2.8 (0.45 dex).

Metallicity variations cannot explain a reduction in \CCOO\ emission, as while at smaller metallicities there is more \CCOO\ dark gas, cluster galaxies show similar or elevated metallicity over isolated systems \cp[e.g.,][]{skillman96,gupta18,franchetto20}.
Changes to the UV radiation field due to {reduction} in star formation also are unlikely to cause any differences in \CCOO\ emission. 
VERTICO galaxies show {reduced} SFR surface densities up to $\sim$1\,dex compared to  star-forming galaxies at fixed $\SigMstar$ (Brown et al., in prep); however, in the same regions, we observe reduced \HI\ and molecular gas densities with similar magnitude, which reduces the gas column's ability to self-shield. 
Thus, any changes in the UV field are likely balanced by changes in the gas density \cp{krumholz09}.
{VERTICO galaxies also show similar $R_{21}$ values to normal galaxies from xCOLD GASS \cp{brown21}, suggesting that there is no significant difference in the \CCOO\ excitation that could under-estimate the amount of molecular gas.}
{Additionally, \ct{smith14} and \ct{glover16} found that \CCOO-dark gas is typically diffuse and low surface density, making it prone to turbulent compression and gravitational collapse and thus less likely to exist in the central regions of galaxies that we have studied.}

Further evidence that VERTICO galaxies do not have substantial \CCOO-dark gas comes from studies of the Kennicutt-Schmidt \cp{schmidt59,kennicutt98} star formation law. 
If the \CCOO\ emission line underestimated the total molecular gas mass, then galaxies would deviate from the Kennicutt-Schmidt law and appear to have elevated SFR at fixed molecular gas mass.
However, VERTICO galaxies show similar galaxy-to-galaxy variation in their resolved Kennicutt-Schmidt relations to what is observed in  star-forming galaxies \cp[][]{vollmer12,jimenez-donaire22}. 
This consistency with unperturbed systems implies that the physics of star formation, and by extension the ISM physical conditions giving rise to the \CCOO\ emission, do not show sufficient difference between VERTICO galaxies and the control sample to explain our results. 
Thus, we need to invoke a process that is unrelated to the ISM physical conditions and can reduce molecular gas content on timescales similar to the \HI\ gas. 
Direct stripping of the molecular gas meets this requirement. 

There is evidence that not all molecular gas in galaxies resides in bound GMCs, with $20-50$ per cent \cp{polk88,wilson94,pety13,roman-duval16,chevance20,saintonge22} existing in a more diffuse phase or a molecular thick disc that has velocity dispersion similar to the \HI\ \cp[e.g.][]{caldu-primo13,caldu-primo15,mogotsi16}, making it much easier to strip than GMC gas.
Our data cannot differentiate between diffuse and dense gas; however, the 64 per cent reduction {(\S\ref{subsec:RMGMS})}  in the typical molecular gas content {within $\RHI$} observed in our large def-\HI\ galaxies is {greater} than the `typical' diffuse gas fractions {mentioned above}.
This is interesting, as there is also evidence that GMCs can be disrupted by the ICM.
\ct{cramer20} found that only 12 per cent of the molecular gas on the leading edge of NGC 4402 exists in GMCs that {have} the survived {the} ongoing stripping, while the remaining gas exists in a trailing plume with lower {surface} density. 
The authors suggest that this large fraction of lower surface density molecular gas implies that an indirect stripping process, by which GMCs are converted to lower {surface} density molecular gas or \HI\ before being directly stripped, could be taking place.

This interpretation is also supported by simulations, which have shown that the multi-phase structure of the ISM allows stripping to affect gas within the truncation radius \cp[e.g., ][]{quilis00,tonnesen09}. 
Further, momentum-transfer from the ICM to the ISM through turbulent mixing can effectively ablate the dense, cold ISM into warmer phases \cp[e.g.,][]{gronke18,sparre19} and accelerate it out of the galaxy \cp{tonnesen21,choi22}.
The typical ICM pressures expected for Virgo galaxies are within the ranges used in these simulations \cp[e.g.,][]{vollmer01,roediger07,chung07,abramson11}, making stripping of the molecular gas the most feasible explanation for our observed molecular gas reduction. 
However, future work with higher spatial resolution observations {is} needed to allow  quantification of how the diffuse and dense phases of the ISM are impacted by the Virgo environment, and search for signatures of direct stripping of the molecular gas.

\section{Summary} \label{sec:concl}
In this paper, we used spatially resolved scaling relations between $\SigMstar$, $\SigHI$, and $\SigMG$ to study how {cluster} {environmental processes} impact the cold gas content within the optical {extent} of {massive ($\lgMstarMsun\geq10$)} galaxies on 1.2-kpc scales.
We presented a new view of how the \HI\ gas distribution changes with increasing \HI\ deficiency, and we have shown, {statistically}, how {environmental processes} impact the {spatial distribution of the} molecular gas, the distribution of $\SigMG$, {and molecular-to-atomic gas mass ratio} using observations from the VERTICO survey \cp{brown21}.
Our main results are as follows:
\begin{itemize}
    \item While the stripping of gas from galaxies impacts the \HI\ gas at large radii, it has the {additional} effect of lowering the average \HI\ {surface} density of the remaining gas within the \HI-stripping radius, $\RHI$. 
    This reduction in {surface} density increases as the gas removal proceeds inward until the densest \HI\ regions can no longer form (Figs. \ref{fig:RHIMSpanels}, \ref{fig:HI_Mstarbins}, \ref{fig:HI_intr}). 
    \item Some galaxies show co-spatial truncation of their \HI\ and molecular gas (\fig{RMGMSpanels}), indicating that {environmental processes} can effectively remove both phases on similar timescales {in the same spatial region}. 
    \item Overall, the molecular gas is less affected than the \HI, and {we find evidence that environmental mechanisms} first {reduce} the lower {surface} density molecular gas that exists preferentially at {{lower} $\SigMstar$ (and thus typically larger radii)} in galaxies (\fig{MG_Mstarbins}).
    However, {we find that} once the removal of \HI\ proceeds sufficiently far within the optical disc, the average {surface} density of the remaining molecular gas {within $\RHI$} is {reduced} (\fig{MG_intr}). 
    \item At the spatial scales we are sensitive to, we see no evidence for environment{-driven} elevation of the \HI\ or molecular gas {surface} density outside the ranges observed in  star-forming galaxies (Figs. \ref{fig:HI_Mstarbins}, \ref{fig:MG_Mstarbins}). 
    \item VERTICO galaxies have systematically larger molecular-to-atomic gas mass ratios in individual regions compared to the control sample, {but smaller values than expected from the removal of \HI\ alone (Figs. \ref{fig:MGHIr_MPP}, \ref{fig:MGHIr_CG}).}
    Using proxies for the physical conditions in the ISM, we showed that changes to the physical state of the gas could not explain the {observed} molecular-to-atomic gas mass ratios (Figs. \ref{fig:MGHIr_MPPbins}, \ref{fig:MGHIr_CGbins}).
    Thus, ablation and/or stripping of the molecular gas on a similar timescale as the \HI\ {in a given 1.2-kpc region} is the best explanation for the reduction of molecular gas content in VERTICO galaxies.
    \end{itemize}

{Our results agree with, and extend upon, previous literature focused on gas stripping and environmental effects in cluster galaxies \cp{cortese21,boselli22}.}
{Combined with the recent, complementary studies from the VERTICO team \cp{brown21,villanueva22,zabel22}, a coherent evolutionary scenario for Virgo cluster galaxies begins to emerge.}
{Compared to isolated systems, statistically, Virgo galaxies have normal molecular gas content \cp{boselli14,cortese16}  and elevated molecular-to-atomic gas mass ratios \cp{mok16,villanueva22} when considering their integrated properties only.}
{Subsequently, there is poor correlation between \HI\ and molecular gas deficiency measurements \cp{zabel22}, but this does not mean that the molecular gas is not affected by {environmental processes}.}
{Instead, major changes in the \textit{integrated} molecular gas mass occur only once a significant fraction of the cold gas within the stellar disc has been affected.}

{Our analysis reveals how the molecular gas content of galaxies is affected by  {cluster environmental processes} in a spatially-resolved way.}
{We have shown that the {reduction} of molecular gas begins at the edge of the stellar disc even in moderately \HI-deficient galaxies, while in the highest \HI-deficiency systems the molecular gas becomes significantly {reduced} across the disc.}
{While this result has been shown using integrated measurements and radial profiles \cp{villanueva22,zabel22}, our approach allows us to quantify this effect as a function of $\SigMstar$ with exquisite detail, and has allowed us to disentangle between environmentally-driven and secular/internal processes.}

{Interestingly, previous works have found evidence for the compression of molecular gas due to the ICM--ISM interaction \cp[e.g.,][]{lee17,moretti20,moretti20a}, particularly in the early stages of infall.}
{We find no evidence for elevated \HI\ or molecular gas surface densities, suggesting that in our sample, which spans a larger (and later) range of infall stages and at the 1.2-kpc scales we are sensitive to, environment-driven elevation of cold gas content is not significant.} 

Our results also emphasise that the gas removal process is more complex than a simple outside-in stripping scenario, {and it is clear that the {environmental processes} affect both  the \HI\ and molecular gas within the \HI\ truncation radius \cp[see also][]{lee22}, which is typically assumed to be the boundary of where stripping is efficient.}
Even at moderate \HI\ deficiency, we find that the average {surface} density of \HI\ remaining within the disc is {reduced}, indicating that the environment does not just affect the \HI\ gas in the outskirts. 
This result can be understood as the multi-phase ISM enabling the ICM to stream through low-density regions and affect the gas across the disc \cp[e.g.,][]{tonnesen09,lee20}.
However, we have also shown that the molecular gas is affected similarly, although it requires a stronger ICM--ISM interaction (as traced by more \HI\ lost). 
Thus, {the} ICM can ablate and strip gas across the entire discs of galaxies and does not discriminate which phase is affected so long as the interaction is strong enough \cp[e.g.,][]{tonnesen21,choi22}.
Indeed, this agrees well with observations of larger molecular gas masses in the cometary tails of stripped galaxies when the associated $\HA$ emission is brighter or the stripping is more advanced \cp{moretti18}. 
In these cases, it is likely that the stronger stripping has led to a larger mass-loading of the outflowing gas.
{Future work with larger samples of field and cluster galaxies, but in particular, higher resolution observations that can separate dense and diffuse \CCOO\ emission will be key to understanding the details of gas suppression in cluster galaxies.}

Last, these results have interesting implications for the quenching of star formation in cluster galaxies. 
In particular, the reduction of molecular gas within the inner regions of galaxies will cause star formation to shut down earlier, and with a different distribution, than just radial truncation. 
Comparison to deep $\HA$ observations of VERTICO galaxies from the VESTIGE \cp[A Virgo Environmental Survey Tracing Ionised Gas Emission,][]{boselli18} survey would enable the study of changes to recent, massive star formation in regions where molecular gas is {reduced}. 
Further, future observations with sensitive optical integral-field spectroscopic instruments such as MUSE \cp[Multi Unit Spectroscopic Explorer,][]{bacon17} will allow the study of the physics of galaxy quenching in unprecedented detail, including the additional impact of feedback from active galactic nuclei \cp[e.g.,][]{boselli16,george19,peluso22}. 
In particular, employing spectro-photometric fitting techniques \cp[e.g.,][]{fossati18} will enable high spatial resolution studies of star formation and star-formation history reconstruction across the discs of Virgo cluster galaxies. 
These results can then be connected to cold gas {reduction} on sub-kpc scales where ISM physics operates.

\begin{acknowledgement}
We thank the referee for their useful comments that improved this paper, and Aaron Robotham for useful conversations. 
We wish to acknowledge the custodians of the land on which much of this work was undertaken, the Wadjuk (Perth region) people of the Nyoongar nation, and their Elders past, present and future.
Part of this work was conducted on the unceded territory of the Lekwungen and Coast Salish peoples. 
We acknowledge and respect the Songhees, Esquimalt, WS\'{A}NE\'{C} and T'Sou-ke Nations whose historical relationships with the land continue to this day.
L.C. and A.B.W acknowledge support from the Australian Research Council Discovery Project  funding scheme (DP210100337). 
L.C. is the recipient of an Australian Research Council Future Fellowship (FT180100066) funded by the Australian Government.
TB acknowledges support from the National Research Council of Canada via the Plaskett Fellowship of the Dominion Astrophysical Observatory.
C.D.W. acknowledges support from the Natural Sciences and Engineering Research Council of Canada and the Canada Research Chairs program.
The financial assistance of the National Research Foundation (NRF) towards this research is hereby acknowledged by N.Z. Opinions expressed and conclusions arrived at, are those of the author and are not necessarily to be attributed to the NRF.
N.Z. is supported through the South African Research Chairs Initiative of the Department of Science and Technology and National Research Foundation.
I.D.R acknowledges support from the ERC Starting Grant Cluster Web 804208.
T.A.D. acknowledges support from the UK Science and Technology Facilities Council through grants ST/S00033X/1 and ST/W000830/1.
A.C. acknowledges the support from the National Research Foundation grant No. 2018R1D1A1B07048314.
A.R.H.S. acknowledges receipt of the Jim Buckee Fellowship at ICRAR-UWA.
K.S. and L.C.P. acknowledge support from the Natural Sciences and Engineering Research Council of Canada (NSERC).
Y.M.B. gratefully acknowledges funding from the Netherlands Organization for Scientific Research (NWO) through Veni grant number 639.041.751.
V.V. acknowledges support from the scholarship ANID-FULBRIGHT BIO 2016 - 56160020 and funding from NRAO Student Observing Support (SOS) - SOSPA7-014. 
A.D.B. and V.V., acknowledge partial support from NSF-AST2108140.
BL acknowledges the support from the Korea Astronomy and Space Science Institute grant funded by the Korea government (MSIT) (Project No. 2022-1-840-05).
Parts of this research were supported by the Australian Research Council Centre of Excellence for All Sky Astrophysics in 3 Dimensions (ASTRO 3D), through project number CE170100013.
This work was carried out as part of the VERTICO collaboration. 
This paper makes use of the following ALMA data: \\
ADS/JAO.ALMA \href{https://almascience.nrao.edu/asax/?result\_view=observation\&projectCode=\%222019.1.00763.L\%22}{\#2019.1.00763.L},\\ 
ADS/JAO.ALMA \href{https://almascience.nrao.edu/asax/?result\_view=observation\&projectCode=\%222017.1.00886.L\%22}{\#2017.1.00886.L},\\ 
ADS/JAO.ALMA \href{https://almascience.nrao.edu/asax/?result\_view=observation\&projectCode=\%222016.1.00912.S\%22}{\#2016.1.00912.S},\\ 
ADS/JAO.ALMA \href{https://almascience.nrao.edu/asax/?result\_view=observation\&projectCode=\%222015.1.00956.S\%22}{\#2015.1.00956.S}.\\
ALMA is a partnership of ESO (representing its member states), NSF (USA) and NINS (Japan), together with NRC (Canada), MOST and ASIAA (Taiwan), and KASI (Republic of Korea), in cooperation with the Republic of Chile. The Joint ALMA Observatory is operated by ESO, AUI/NRAO and NAOJ. The National Radio Astronomy Observatory is a facility of the National Science Foundation operated under cooperative agreement by Associated Universities, Inc.
\end{acknowledgement}

\printendnotes

\bibliography{VERTICO_RMSs}

\appendix

\section{Combined and individual $\SigMGHI$ relations} \label{sec:app1}

\begin{figure*}
    \centering
    \includegraphics[width=0.49\textwidth]{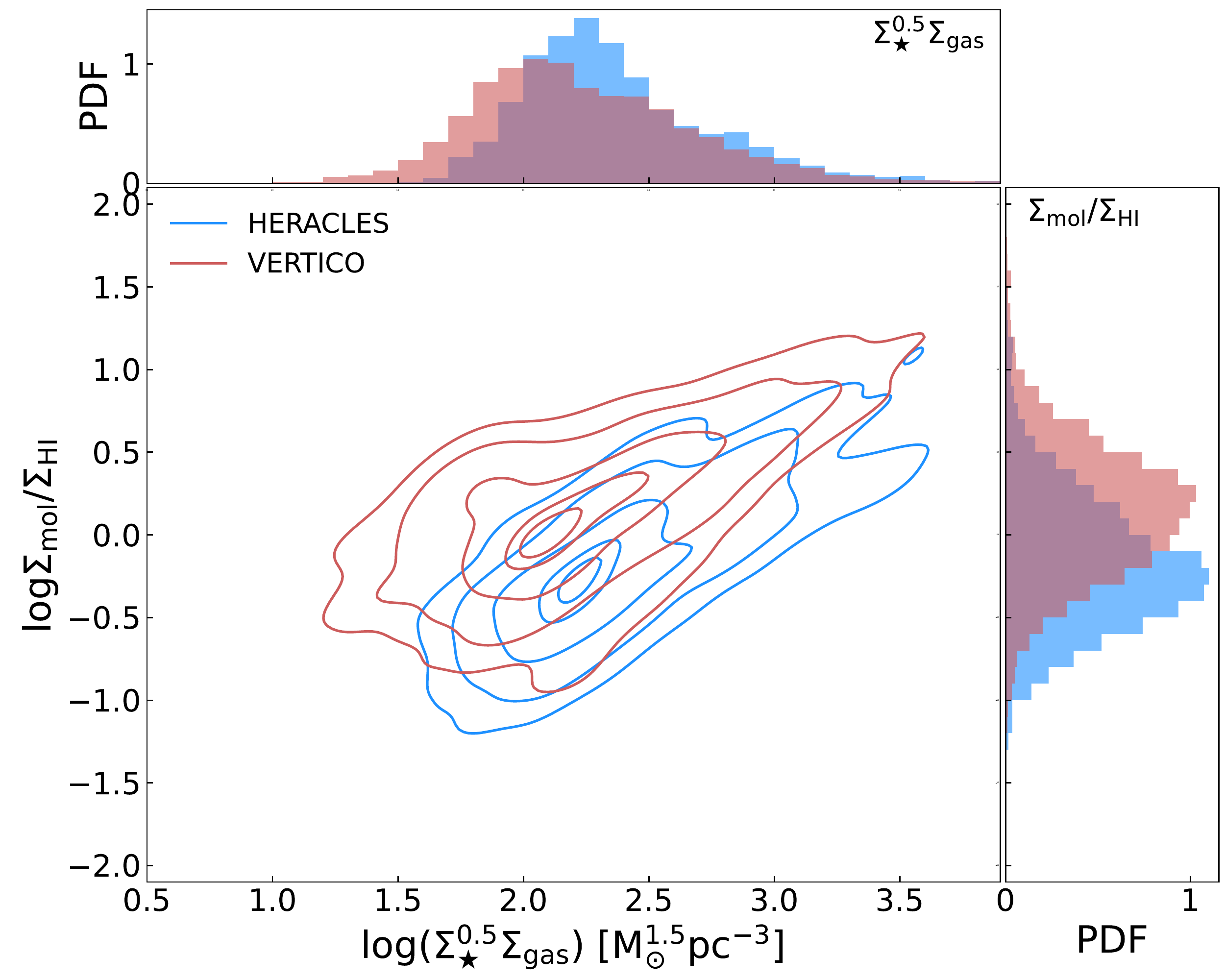}
      \includegraphics[width=0.49\textwidth]{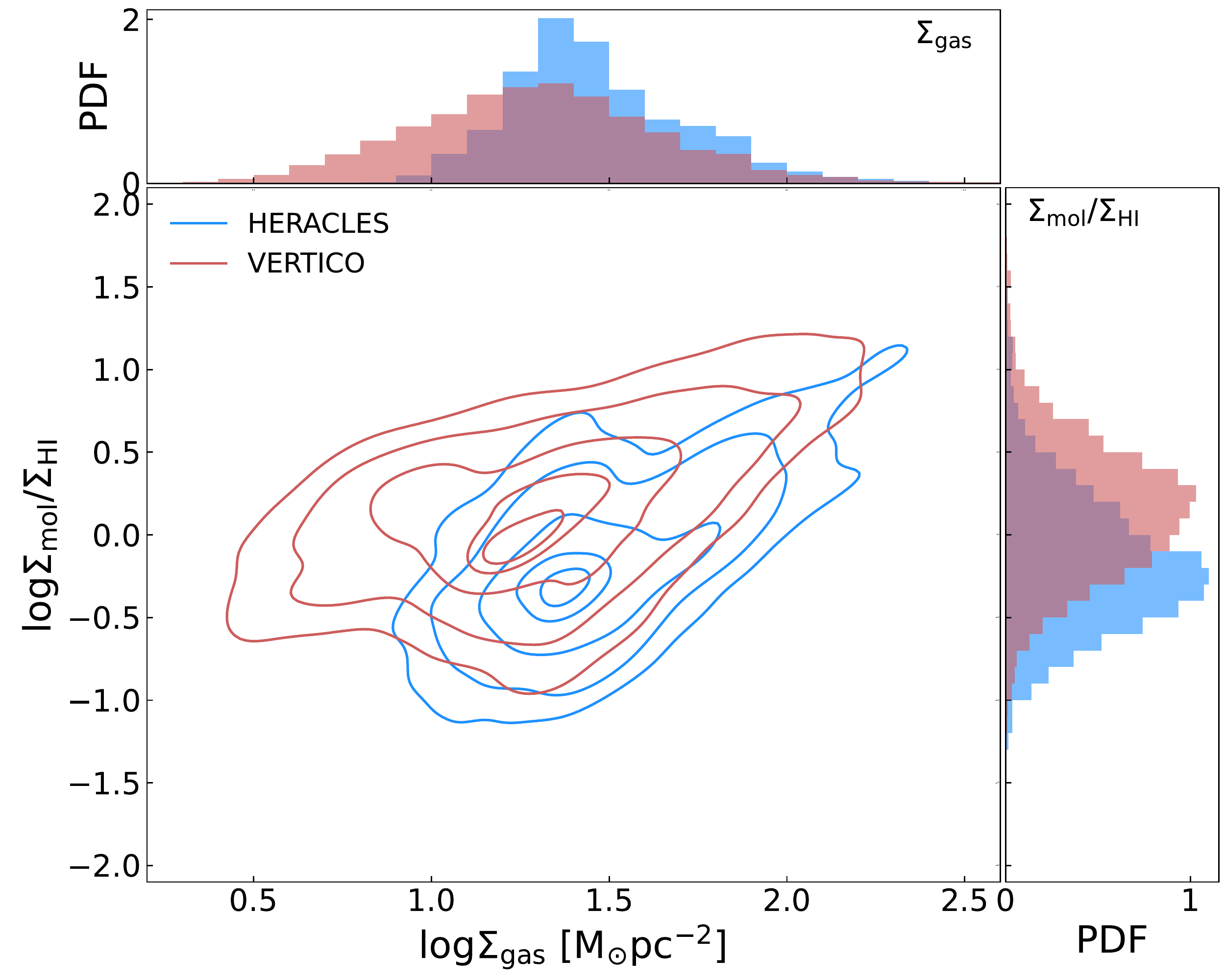}
    \caption{
    Combined versions of the spatially resolved $\SigMGHI$ relations as a function of two proxies for the ISM physical conditions.
    The left panel shows $\SigMGHI$ vs the midplane-pressure proxy $\log \SigMstar^{0.5}\SigGas$, and the right panel vs the total gas surface density $\SigGas$. 
    Contours levels for both samples enclose the  5, 16, 50, 84, and 95 per cent of the sample.
    Density-normalised histograms of $\lgSigMGHI$, $\log \SigMstar^{0.5}\SigGas$, and $\lgSigGas$ are shown opposite their respective axes.
    }
    \label{fig:combRMPPr_RMFERr}
\end{figure*}

\begin{figure*}
    \centering
    \includegraphics[width=\textwidth]{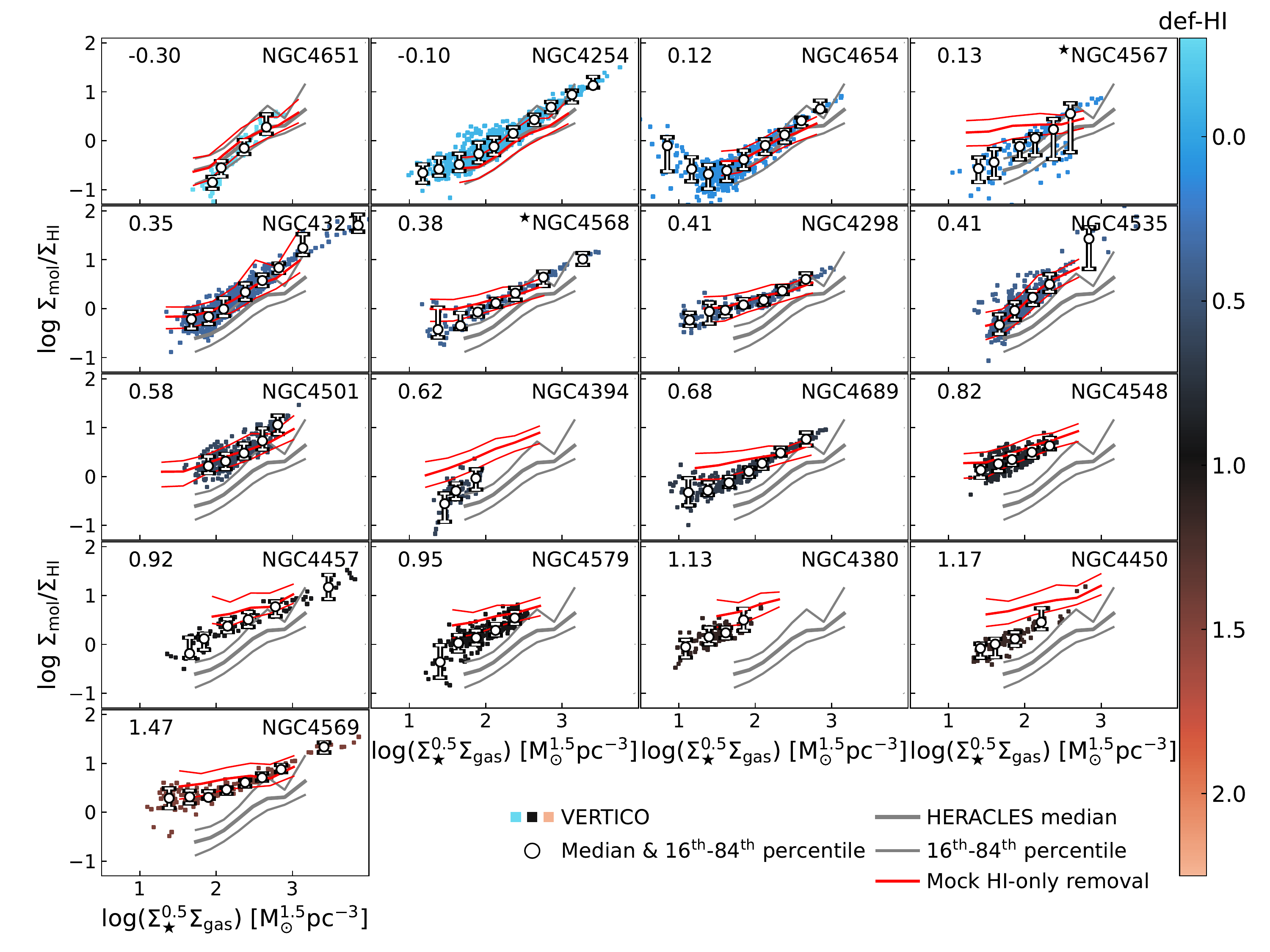}
    \caption{
    Individual $\lgSigMGHI - \log \SigMstar^{0.5}\SigGas$ relations for VERTICO galaxies, with the galaxy name given in the top-right {of each panel}.
    The panels are ordered in rows by def-\HI\ {(noted in the top-left corner)}, which also sets the colour of each panel, and a colour bar is displayed on the right-hand side. 
    The median relation for each galaxy computed in 0.25\, dex $\lgSigMstar$ bins are shown with large white markers, {and error bars show the 16$^\mathrm{th}$ and 84$^\mathrm{th}$ percentiles of the $\lgSigMGHI$ distribution in each bin.}.
    The HERACLES control sample is shown with thick (median) and thin (16$^\mathrm{th}$ and 84$^\mathrm{th}$ percentiles).
    The shift in the HERACLES median that would be expected from the observed {reduction} in $\SigHI$ at fixed $\SigMstar$ for each galaxy is shown with a solid red line. 
    The two galaxies marked with stars are interacting, and their \HI\ observations are partially confused.
    }
    \label{fig:RMPPrpanels}
\end{figure*}

\begin{figure*}
    \centering
    \includegraphics[width=\textwidth]{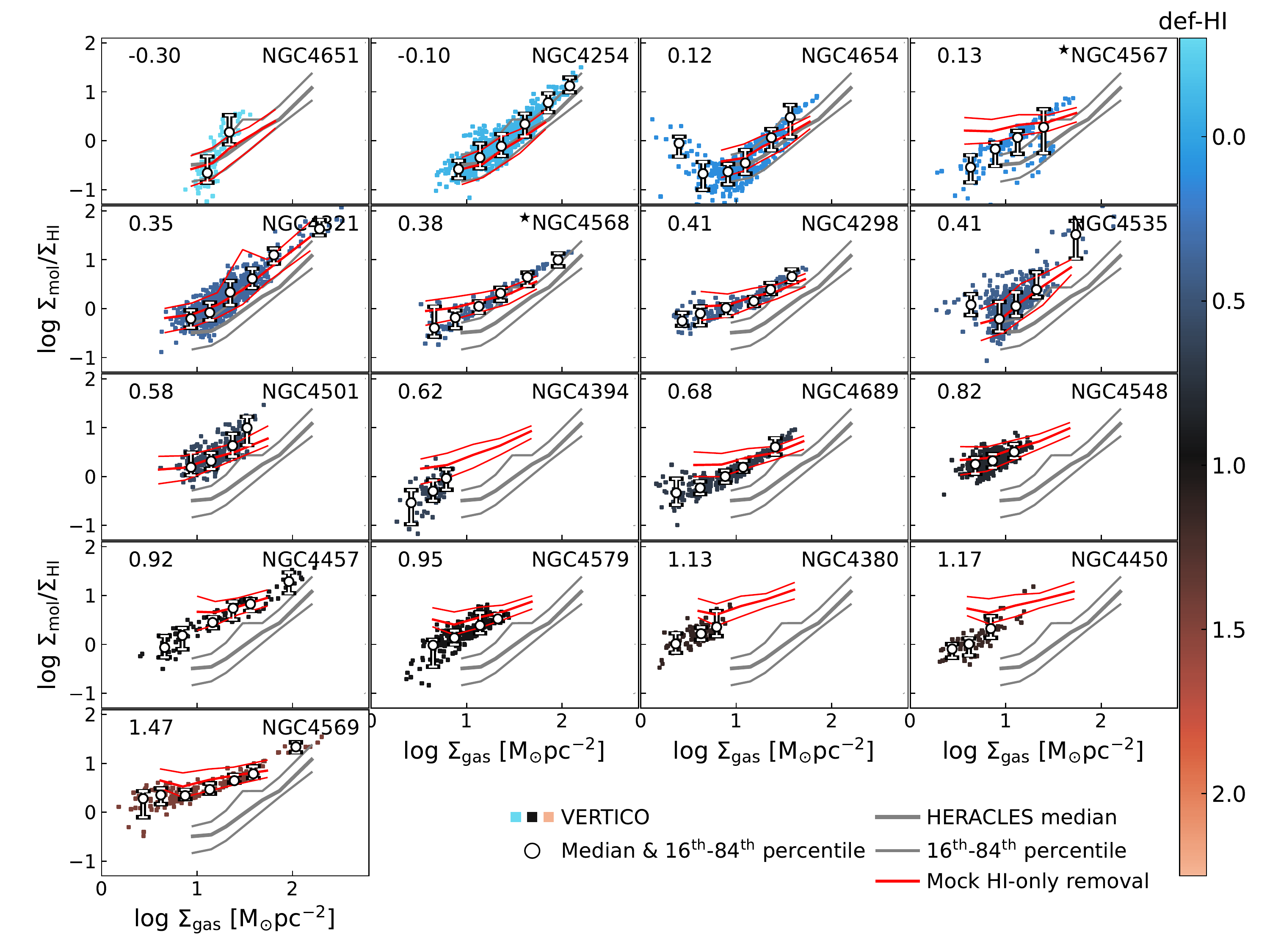}
    \caption{
    The same as \fig{RMPPrpanels}, but for the individual $\lgSigMGHI -\lgSigGas$ relations.
    }
    \label{fig:RMFErpanels}
\end{figure*}

\end{document}